\documentclass[aps,prd,twocolumn,floatfix]{revtex4-2}
\usepackage{eurosym}
\usepackage{amsfonts}
\usepackage{amssymb}
\usepackage{amsmath}
\usepackage{dsfont}
 \usepackage[dvipsnames]{xcolor}
\usepackage{float}
\usepackage{accents}
\usepackage{graphicx}
\usepackage{tikz}
\usepackage{tkz-graph}
\usepackage[latin1]{inputenc}
\usepackage{epstopdf}
\usepackage{soul}
\usepackage{ulem}
\usepackage[colorlinks=true,pdfstartview=FitV,linkcolor=blue,citecolor=blue,urlcolor=blue,breaklinks=true]{hyperref}

\usepackage{orcidlink}

\RequirePackage{color}

\begin{document}

\title{Self-dual compactons in the gauged restricted baby Skyrme model {in the presence of an external magnetic field}}

\author{N. H. Gonzalez-Gutierrez$^a$ \orcidlink{0000-0001-9120-0179}}
\email{neyver.hgg@discente.ufma.br}\email{neyver.hgg@gmail.com}
\author{Rodolfo Casana$^{a,b}$ \orcidlink{0000-0003-1461-3038}}
\email{rodolfo.casana@ufma.br}\email{rodolfo.casana@gmail.com}
\author{André C. Santos$^{a,c}$ \orcidlink{0000-0002-1413-8938}}
\email{andresantos@ccn.uespi.br}\email{andre$\_$cavs@hotmail.com}
\affiliation{$^a$Programa de Pós-graduação em Física, Universidade Federal do Maranhão, 65080-805, São Luís, Maranhão, Brazil.}
\affiliation{$^b$Departamento de Física, Universidade Federal do Maranhão, 65080-805, São Luís, Maranhão, Brazil.}
\affiliation{$^c$Centro de Ciências da Natureza, Universidade Estadual do Piauí, 64002-150, Teresina, Piauí, Brazil.}

\begin{abstract}

We investigate the existence of compact self-dual solitons in the restricted gauged baby Skyrme model in the presence {of an external magnetic field.} The consistent implementation of the Bogomol'nyi-Prasad-Sommerfield (BPS) formalism depends on the relative size between the compacton and the effective region occupied by the {external} magnetic field. {To address this issue, we consider {  two} {scenarios: in} the first, the external magnetic field is confined within the compacton, effectively playing the role of a magnetic {  impurity; in the second scenario,} the external magnetic field fully encircles} {the compacton. For} both cases, the approach has enabled us to set the self-dual potential, achieve the Bogomol'ny bound for the energy, and establish the self-dual or BPS equations whose solutions saturate such a bound. {We next focused on obtaining radially symmetric compactons by solving the BPS system using two functions to describe the external magnetic {  field, a step-type function and a Gaussian function.}} After solving the BPS system numerically, we depicted the resulting field profiles and highlighted the effects induced on the compacton's size, field profiles, magnetic field, and magnetic flux.

\end{abstract}

\maketitle

\section{Introduction}

Effective field theories play a crucial role in contemporary physics, especially when they offer insights into specific system properties that remain challenging to extract starting from the corresponding high-energy model. One of the most fundamental still unsolved issues in modern theoretical physics is searching for an effective low-energy model derived entirely from Quantum Chromodynamics (QCD). In this context, the  (1+3)-dimensional Skyrme model \cite{skyrme} emerges as a promissory effective theory for describing such a regimen  \cite{zahed, adkinsx1, adkinsx2, Mantonx1, Mantonx2, Mantonx3, Mantonx4, Mantonx5, max1,max2, max3, holtx1, holtx2, holtx3, sutcliffe}. This nonlinear model identifies the baryons as collective excitations  (i.e., topological solitons called skyrmions)  of the fundamental QCD fields. However, one of the main problems of the Skyrme model is providing binding energies that are one order of magnitude larger than the respective experimental nuclear data \cite{sutcliffe}. In the face of such a problem, searching for a model embodying a Bogomol'nyi-Prasad-Sommerfield (BPS) structure  \cite{Bogomolnyix1, Bogomolnyix2} is a possible starting point for circumventing this problem. Such a class of theories possesses an energy lower-bound (the Bogomol'ny bound) of topological character and stable solitons  (solutions of the so-called BPS equations) that saturate such a bound. Since the masses of atomic nuclei are nearly linear on the baryon charge, investigating (near) BPS skyrmions becomes a convenient approach because the Bogomol'ny bound depends linearly on the topological (baryon) charge,  {resulting in null classical binding energies. Moreover, adding to the Lagrangian density quantum corrections and small contributions of additional terms, we can achieve realistic small values for nuclear binding energies} \cite{Bonenfantx1, Sutcliffe02, Bonenfantx2, Gudnasonx1, Gudnasonx2}.

{It is worthwhile to highlight that although initially the Skyrme model was} developed to describe the internal structure of atomic nuclei, it afterward inspired the concept of magnetic skyrmions {(a topological soliton describing the magnetization inside a medium) that characterizes} localized finite energy {configurations appearing} in magnetic materials. {Such topological structures} arise due to the chiral interaction among spins present in magnetic materials without inversion symmetry {that becomes} stabilized by the Dzyaloshinskii-Moriya interaction (DMI) \cite{Dzyaloshinsky, Moriya}, which prevents the collapse of the magnetization vector field into a magnetic singularity. {The so-called magnetic skyrmions are investigated along a wide breadth of physical phenomena, for instance,} including 2D electron gases \cite{Brey}, spinor Bose-Einstein condensates \cite{Choi}, and their prospective applications in spintronics \cite{Fert}, which have the promise to shape the future landscape of digital electronics.

{Meanwhile, the study of the $(1+2)$-dimensional version of the Skyrme model{, known as the baby Skyrme model \cite{piettex1, piettex2, piettex3, piettex4, gisiger1996},} has served as a laboratory for understanding many aspects of Skyrme's original model. In this sense, skyrmions have garnered considerable attention from the scientific community, as they are either employed in or emerge in the description of several physical systems. Among them,} we can mention the topological quantum Hall effect \cite{Balram}, chiral nematic liquid crystals \cite{Fukuda}, superconductors \cite{Zyuzin}, brane cosmology \cite{Delsate}, and magnetic materials \cite{Yu}, including recent investigations with DMI \cite{Schroers}.

{The (ungauged) standard baby Skyrme model describes stable solitons but does not possess a BPS structure. Nevertheless, the so-called restricted baby Skyrme model \cite{gisiger} admits BPS configurations \cite{adamx1, adamx2, adamx3}.} {Subsequently, the authors in Ref. \cite{adam2} initiated the search for BPS configurations in the gauged restricted baby Skyrme {model, which} carry only magnetic flux (see also Refs. \cite{adam5, Casana065009}), including the study of topological structures known as compactons. Such solutions, {  also referred as} compact Skyrmions in this context, are solitons whose profiles reach the vacuum value at a finite distance.} Similarly, recent studies have shown that BPS skyrmions can carry both magnetic flux and electric charge \cite{CAdamcs, Casana045022, Casana045018}. Moreover, Refs. \cite{J.Andrade, Casana220906309} have performed the search for BPS solitons in the gauged restricted baby Skyrme model immersed in a magnetic medium. {Further,} Refs. \cite{susy1, susy2, susy3, queiruga} have investigated aspects related to supersymmetry and the BPS states, whereas Refs. \cite{gravity1, gravity2} have analyzed the presence of BPS skyrmions in gravitational theories.

Recently, the study of topological structures in the presence of impurities has been gaining increasing attention. Including an impurity (a non-dynamical background field) in a physical system is equivalent to embedding the field theory describing it within a nontrivial medium. The impurities may arise or emerge inside a physical system from local inhomogeneities or global dependencies on external parameters. Consequently, physical environments such as condensed matter \cite{Shapoval}, Bose-Einstein condensates \cite{Tung, Griffin}, inclusive in neutron stars \cite{Anderson, Wlazowski} have been explored to study impurity's effects on topological states. In the framework of self-dual systems, a BPS model coupled to an impurity may, in general,  lose its BPS structure completely, i.e., the solitons solutions do not saturate the correspondent BPS bound. {However, in Refs. \cite{AdamW, AdamRW, AdamKQ}, the authors have shown that it is possible to construct an impurity model preserving half of the BPS property, resulting in so-called half-BPS solutions, such as revealed by the supersymmetric extensions of impurity models defined in (1+1)- and (1+2)-dimensions \cite{SuperAdam}}. Moreover, in the literature, there are well-established studies of BPS systems emerging from gauge field theories enlarged by impurities. For example, in kink-like solutions \cite{D.Baz}, in the Maxwell-Higgs scenario \cite{Tong}, Chern-Simons-Higgs model \cite{Han}, in gauged CP(2) case  \cite{daHora}, the interaction between a moving Maxwell-Higgs vortex and a static magnetic impurity \cite{Ashcroft}.

{A natural question in this context is whether the gauged BPS baby Skyrme model also supports a self-dual structure capable of generating compactons when a magnetic impurity or an external magnetic field is present. Therefore, {  this} manuscript proposes developing such a BPS structure. Our approach enables the investigation of the magnetic impurity's or external magnetic fields' effects on the compacton's formation, which is associated with the spatial extent of the applied external field.} For this purpose, we have organized our results as follows: In Sec. \ref{Sec02}, we introduce the gauged baby Skyrme model in the presence of {an external magnetic field.} {
Sec. III is devoted to implementing the BPS formalism for the model by considering two scenarios: first, the external magnetic field is {wholly held} within the compacton, playing the role of a magnetic impurity (internal scenario); second, the compacton is contained entirely inside the external magnetic field, i.e., the field extends beyond of the region occupied by compact skyrmion (extended scenario).} This way, we successfully achieve the energy lower bound (the Bogomol'ny bound) and the corresponding self-dual or BPS equations whose solutions saturate this bound. Besides, one presents the ansatz for attaining radially symmetric solitons. In Sec. \ref{Sec04}, we define the form of the superpotential capable of generating compactons in the presence of {an external magnetic field}. {Next, we explore the effects of the external magnetic field by setting two functions:} the first one involves a step-type function, and the second case incorporates a Gaussian function. Finally, we highlight our results and provide our perspectives in Sec. \ref{Sec05}.

\section{The model\label{Sec02}}

We introduce our effective model, which describes the restricted gauge
Skyrme model \cite{adam2} in the presence of a magnetic source term, by
taking the Lagrangian%
\begin{equation}
L=E_{0}\int d^{2}\mathbf{x}\,{\mathcal{L}}\text{,}  \label{EQ1}
\end{equation}%
with $E_{0}$ defining the energy scale (which we set $E_{0}=1$ in the
remainder of the manuscript), and the Lagrangian density given by
\begin{equation}
\mathcal{L}=-\frac{1}{4g^{2}}F_{\mu \nu }^{2}-\frac{\lambda ^{2}}{4}(D_{\mu}
\vec{\phi}\times D_{\nu }\vec{\phi})^{2}+\Delta B-U(\phi _{n},\Delta ).  \label{EQ5}
\end{equation}%
The coupling between the gauge field and the Skyrme field is given through
the covariant derivative
\begin{equation}
D_{\mu }\vec{\phi}=\partial _{\mu }\vec{\phi}+A_{\mu }(\hat{n}\times \vec{%
\phi})\text{,}  \label{EQ3}
\end{equation}%
being $\vec{\phi}$ the Skyrme field, which is a three-component vector of
scalar fields $\vec{\phi}=\left( \phi _{1},\phi _{2},\phi _{3}\right) $
obeying $\vec{\phi}\cdot \vec{\phi}=1$, and hence describing the unit sphere
$\mathbb{S}^{2}$. The unitary vector $\hat{n}$ gives a preferential
direction in the internal space $\mathbb{S}^{2}$.

The first term in Eq. (\ref{EQ5}) is the Maxwell term, with $F_{\mu\nu}= \partial _{\mu }A_{\nu }-\partial _{\nu }A_{\mu }$, being $A_{\mu }$ the
Abelian gauge field and $g$ the \ electromagnetic coupling constant. The
second contribution is the Skyrme term, where $\lambda $ is the
corresponding coupling constant. {In the third term, we have introduced the external magnetic field through the function} $\Delta =\Delta ({x})$, which couples linearly to the magnetic field $B=F_{12}$. Lastly, the function $U(\phi _{n},\Delta )$ stands for a potential that also is a function of the magnetic impurity, where $\phi _{n}\equiv \hat{n}\cdot \vec{\phi}$. Moreover, both coupling constants are assumed non-negative, being $g$ and $\lambda $ with mass
dimension $1$ and $-1$ respectively; the gauge field has mass dimension $1$,
and the Skyrme field is dimensionless.

The Euler-Lagrange equations obtained from Eq. (\ref{EQ5}) are
\begin{equation}
\partial _{\nu }F^{\nu \mu }-g^{2}\left( \delta _{2}^{\mu }\partial
_{1}-\delta _{1}^{\mu }\partial _{2}\right) \Delta =g^{2}j^{\mu }\text{,}
\label{EQ6}
\end{equation}%
\begin{equation}
D_{\mu }\vec{J}^{\mu }+\frac{\partial V}{\partial \phi _{n}}(\hat{n}\times
\vec{\phi})=0\text{,}  \label{EQ7}
\end{equation}%
where $j^{\mu }=\hat{n}\cdot \vec{J}^{\mu }$ is the conserved current
density, with $\vec{J}^{\mu }$ given by
\begin{equation}
\vec{J}^{\mu }=\lambda ^{2}[\vec{\phi}\cdot (D^{\mu }\vec{\phi}\times
D^{\rho }\vec{\phi})]D_{\rho }\vec{\phi}\text{.}  \label{EQ8}
\end{equation}

In this study, we focus on time-independent solutions and hence, from (\ref%
{EQ6}), we obtain the respective Gauss's law
\begin{equation}
\partial _{i}\left( \partial _{i}A_{0}\right) =g^{2}\lambda ^{2}A_{0}(\hat{n}%
\cdot \partial _{i}\vec{\phi})^{2}\text{.}  \label{EQ9}
\end{equation}%
{Here, we observe that Gauss's law is identically satisfied by the gauge condition $A_{0}=0$, meaning that the resulting configurations carry only magnetic flux.} Furthermore, also from Eq. (\ref{EQ6}), we can obtain Ampère's law
\begin{equation}
\partial _{i}(B-g^{2}\Delta )+g^{2}\lambda ^{2}(\hat{n}\cdot \partial _{i}%
\vec{\phi})Q=0\text{,}  \label{EQ10}
\end{equation}%
where already we have considered $A_{0}=0$. The magnetic field is $%
B=F_{12}=\epsilon _{ij}\partial _{i}A_{j}$ and the quantity $Q$ defined as%
\begin{equation}
Q=\vec{\phi}\cdot (D_{1}\vec{\phi}\times D_{2}\vec{\phi})=q+\epsilon
_{ij}A_{i}(\hat{n}\cdot \partial _{j}\vec{\phi})\text{,}  \label{Q}
\end{equation}%
being%
\begin{equation}
q=\frac{1}{2}\epsilon _{ij}\vec{\phi}\cdot (\partial _{i}\vec{\phi}\times
\partial _{j}\vec{\phi})\text{,}  \label{q}
\end{equation}%
a quantity related to the topological charge or topological degree (or
winding number) of the Skyrme field, which is given%
\begin{equation}
\deg [\vec{\phi}]=-\frac{1}{4\pi }\!\int q\,d^{2}\mathbf{x}=N\in \mathds{Z}\setminus \{0\}\text{.}\label{N}
\end{equation}%
Moreover, the stationary version of the Eq. (\ref{EQ7}) of the Skyrme field becomes
\begin{equation}
\lambda ^{2}\epsilon _{ij}D_{i}(QD_{j}\vec{\phi})+\frac{\partial V}{\partial
\phi _{n}}(\hat{n}\times \vec{\phi})=0\text{.}  \label{EqV}
\end{equation}

The following section is devoted to implementing the BPS formalism to investigate the conditions under which the model (\ref{EQ5}) engenders self-dual compact configurations. The formalism allows us to determine the self-dual potential, the Bogomol'ny bound for the total energy, and the self-dual equations whose solutions saturate such a bound.

\section{BPS structure\label{Sec03}}

{{We will} now construct {a} BPS structure {for} the gauged {restricted} baby Skyrme model in the presence of an external magnetic field {that enables us to investigate compact skyrmions}.} For such a purpose, we consider the stationary energy density of the model (\ref{EQ5}) that, under the gauge condition $A_{0}=0$, reads as
\begin{equation}
\mathcal{\varepsilon }=\frac{1}{2g^{2}}B^{2}-\Delta B+\frac{\lambda ^{2}}{2}%
Q^{2}+U\left( \phi _{n},\Delta \right) \text{.}  \label{EQ11}
\end{equation}%

{To ensure the existence {of compact} configurations with finite energy, we define a finite region $\mathcal{R}\subset \mathds{R}^{2}$ as the domain supporting the compactons, with its boundary $\partial \mathcal{R}$ where the fields attain their vacuum values when $\mathbf{x} \in \partial \mathcal{R}$.}
{Consequently, we require that the solutions engendered by the BPS structure possess a null energy density at the compacton's border, i.e., the energy density (\ref{EQ11}) at the boundary $\partial \mathcal{R}$ must obey}
\begin{equation}
	\lim_{\mathbf{x}{\rightarrow }\partial \mathcal{R}}\mathcal{\varepsilon }(%
	\mathbf{x})=0\text{.}  \label{BCEn}
\end{equation}

{Furthermore, the spatial extent of the external magnetic field enables the analysis of two scenarios within the BPS formalism. In the first, named the internal scenario, the external magnetic field is entirely {maintained within} the compact skyrmion, acting as a magnetic impurity. In the {second scenario, called} the extended scenario, the external magnetic field extends beyond the region occupied by the compact skyrmion, {i.e., the compacton size occupies an area wholly located inside the applied external field.} We will address both cases in detail in the subsequent subsections.}

\subsection{BPS structure for the internal scenario \label{internal}}

{We consider the external magnetic field} defined by the function
\begin{equation}
\Delta (\mathbf{x})=\left\{
\begin{array}{c}
\Delta ^{c}(\mathbf{x})\text{ \ \ ,\ \ \ }\mathbf{x}\in \mathcal{R}_{0}, \\[0.2cm]
\text{ \ }0\text{\ \ \ \ \ \ \ ,\ \ \ }\mathbf{x}\notin \mathcal{R}_{0},%
\end{array}%
\right.  \label{Imp0}
\end{equation}%
where $\Delta ^{c}(\mathbf{x})$ is an integrable function with domain into the region $\mathcal{R}_{0}\subset \mathcal{R}$. This way, the {external field} must satisfy the boundary condition%
\begin{equation}
\lim_{\mathbf{x}{\rightarrow }\partial \mathcal{R}}\Delta (\mathbf{x})=0%
\text{,}  \label{Imp1}
\end{equation}
{{guaranteeing that} the field $\Delta (\mathbf{x})$ {remains entirely contained} within the compact skyrmion, {thus playing} the role of a magnetic impurity.}

{Starting from the energy density (\ref{EQ11}) and its vacuum condition (\ref{BCEn}), we {establish the boundary} conditions that fields describing the model must satisfy to ensure compact configurations with finite energy. In this way, the magnetic field, the quantity $Q$, and the potential $U$ must fulfill the following boundary conditions:}
\begin{equation}
\lim_{\mathbf{x}{\rightarrow }\partial \mathcal{R}}Q=0\text{,}\label{Qint}
\end{equation}
\begin{equation}
\lim_{\mathbf{x}{\rightarrow }\partial \mathcal{R}}B=0\text{,}\quad \lim_{%
\mathbf{x}{\rightarrow }\partial \mathcal{R}}U=0\text{.}  \label{bcBQU}
\end{equation}

By taking the integration of the energy density ({\ref{EQ11}), we obtain the system's total energy
\begin{equation}
E=\int_{\mathcal{R}}d^{2}\mathbf{x}\left[ \frac{1}{2g^{2}}B^{2}-\Delta B+%
\frac{\lambda ^{2}}{2}Q^{2}+U\left( \phi _{n},\Delta \right) \right].
\label{EQ14}
\end{equation}%
{In the following, to implement the BPS formalism, we perform some algebraic manipulations allowing us to rewrite (\ref{EQ14}) as}
\begin{eqnarray}
	E &=&\int_{\mathcal{R}}d^{2}\mathbf{x}\left[ \frac{1}{2g^{2}}\left(
	B-g^{2}\Delta \pm \lambda ^{2}g^{2}W\right) ^{2}\right.  \notag \\[0.08in]
	&&~\ \ \ \ \ +\frac{\lambda ^{2}}{2}\left( Q\mp W_{\phi _{n}}\right) ^{2}\pm
	\lambda ^{2}qW_{\phi _{n}}\mp \lambda ^{2}\epsilon _{ij}\partial _{i}\left(
	WA_{j}\right)  \notag \\
	&&\left. ~\ \ \ \ \ +U-\frac{\lambda ^{2}}{2}W_{\phi _{n}}^{2}-\frac{1}{2}%
	g^{2}\left( \lambda ^{2}W\mp \Delta \right) ^{2}\right] \text{,}
	\label{EQ15}
\end{eqnarray}
{where we have introduced the auxiliary function $W(\phi _{n})$ and defined $W_{\phi _{n}}\equiv \partial W/\partial \phi _{n}$.} Further, we have also used $B=\epsilon _{ij}\partial _{i}A_{j}$ and the expression in (\ref{Q}).

{To continue with the BPS implementation, we now require that the potential $U\left( \phi _{n},\Delta \right)$ obeys the equation}
\begin{equation}
U\left( \phi _{n},\Delta \right)=\frac{\lambda ^{2}}{2}W_{\phi _{n}}^{2}+\frac{1}{2}g^{2}\left( \lambda
^{2}W\mp \Delta \right) ^{2}\text{,}  \label{POT}
\end{equation}%
{which provides a relation between the BPS potential, {the magnetic impurity $\Delta (\mathbf{x})$}, and the function  $W(\phi_n)$. Moreover, the} vacuum conditions (\ref{Imp1}) and (\ref{bcBQU}) allow us to establish boundary conditions for the function $W(\phi _{n})$, namely
\begin{equation}
\lim_{\mathbf{x}{\rightarrow }\partial \mathcal{R}}W=0, \quad \lim_{\mathbf{x%
}{\rightarrow }\partial \mathcal{R}}W_{\phi _{n}}=0\text{.}  \label{bcW}
\end{equation}%
Henceforth, we shall call $W(\phi _{n})$ of superpotential due to similarities of the relation between potential and superpotential in supersymmetric field theories.} In what follows, we observe that the contribution of the total derivative $\epsilon _{ij}\partial _{i}\left(WA_{j}\right)$ to the  energy (\ref{EQ15})  {vanishes} due to the boundary condition (\ref{bcW}). {Therefore, assuming the considerations above, we can express} the total energy as
\begin{equation}
E=E_{_{\text{BPS}}}+\bar{E}\text{,}  \label{EQ18}
\end{equation}%
where, by using (\ref{q}), we have defined the Bogomol'ny bound
\begin{equation}
E_{_{\text{BPS}}}=\mp \lambda ^{2}\int_{\mathcal{R}}d^{2}\mathbf{x}W_{\phi
_{n}}\vec{\phi}\cdot (\partial _{1}\vec{\phi}\times \partial _{2}\vec{\phi}%
)\geq 0\text{,}  \label{EQ18B}
\end{equation}%
and the part with the quadratic terms
\begin{eqnarray}
\bar{E} &=&\int_{\mathcal{R}}d^{2}\mathbf{x}\left[ \frac{1}{2g^{2}}\left(
B-g^{2}\Delta \pm \lambda ^{2}g^{2}W\right) ^{2}\right.  \notag \\
&&\text{ \ \ \ \ \ \ \ \ }\left. +\frac{\lambda ^{2}}{2}\left( Q\mp W_{\phi
_{n}}\right) ^{2}\right] \text{.}  \label{EQ18A}
\end{eqnarray}%
Consequently, the total energy (\ref{EQ18}) satisfies the inequality $E\geq
E_{_{\text{BPS}}}$ and the Bogomol'ny bound will be attained when the fields
have configurations such that $\bar{E}=0$, i.e., the bound is saturated when
the following set of first-order differential equations obey%
\begin{equation}
B=g^{2}\Delta \mp \lambda ^{2}g^{2}W\text{,}  \label{EQ20}
\end{equation}%
\begin{equation}
Q=\pm W_{\phi _{n}}\text{,}  \label{EQ21}
\end{equation}%
defining the so-called self-dual or BPS equations of the model. Such a set of equations recovers the Euler-Lagrange equations (\ref{EQ10}) and (\ref{EqV}), associated with the Lagrangian density (\ref{EQ5}) by considering the self-dual potential (\ref{POT}). Furthermore, the solutions of these equations are also classical solutions belonging to an extended supersymmetric model \cite{WittenOlive, Hlousek} whose bosonic sector would be given by the Lagrangian density (\ref{EQ5}). Indeed, the gauged restricted baby Skyrme model supports a $\mathcal{N}=2$  supersymmetric formulation \cite{SuperAdam2}, which is also realizable in the presence of an impurity term \cite{SuperAdam}.

Here, it is essential to point out that if the {external magnetic field} does not meet the conditions (\ref{Imp0}) and (\ref{Imp1}), then the ones established in (\ref{bcBQU}), for the magnetic field and the potential, will not be upheld. Such a situation may occur when {this} {external magnetic field} extends in space beyond the compact boundary $\partial \mathcal{R}$. {In this circumstance, it is necessary to define {suitable} boundary conditions {that are consistent with the new region covered} by the external field, {as will be discussed below.}}

\subsection{BPS structure for the extended scenario\label{extended}}

To study the effects of an {external magnetic field} occupying a region $\mathcal{R}_{0} \subseteq \mathds{R}^{2}$ that is larger than or equal to the region $\mathcal{R}$ covered by compacton ($\mathcal{R}$), we here consider an {external magnetic field} defined by the function $\Delta (\mathbf{x})$ obeying the following condition:
\begin{equation}
\left\vert \int_{\mathcal{R}_{0}}\Delta (\mathbf{x})d^{2}\mathbf{x}%
\right\vert <\infty \text{,}  \label{Imp2}
\end{equation}%
such that $\mathcal{R}\subseteq \mathcal{R}_{0}$.

Now let us establish the boundary conditions to be satisfied by the fields in the presence of the {external magnetic field} defined by Eq. (\ref{Imp2}) to guarantee that the energy density (\ref{EQ11}) be null at compacton's boundary $\partial \mathcal{R}$ such as required by the equation (\ref{BCEn}). Firstly, the boundary condition for the quantity $Q$ remains being%
\begin{equation}
\lim_{\mathbf{x}{\rightarrow }\partial \mathcal{R}}Q=0\text{,} \label{Qext}
\end{equation}%
{while the magnetic field and the potential satisfy, respectively, the
nonnull conditions}
\begin{equation}
\lim_{\mathbf{x}{\rightarrow }\partial \mathcal{R}}B=g^{2}\lim_{\mathbf{x}{\
\rightarrow }\partial \mathcal{R}}\Delta \text{,}\quad \lim_{\mathbf{x}{%
\rightarrow }\partial \mathcal{R}}U=\frac{g^{2}}{2}\lim_{ \mathbf{x}{%
\rightarrow }\partial \mathcal{R}}\Delta ^{2}\text{,}  \label{bcBUext}
\end{equation}%
{where the potential at the frontier has to attain its absolute minimum.
Thus, these new boundary conditions guarantee that energy
density is null when $\textbf{x}\in \partial \mathcal{R}$.}

Once we have fixed the new boundary conditions, the implementation of the BPS
formalism will run similarly to the one developed previously. In this way,
the boundary conditions (\ref{bcW}) established for the superpotential remain the same. Accordingly, the equations describing the BPS configurations hold
their mathematical form, i.e., the BPS potential (\ref{POT}), BPS total
energy (\ref{EQ18B}) and the BPS equations (\ref{EQ20})-(\ref{EQ21}) do not
change their functional or algebraic structure.

\subsection{Magnetic flux \label{MF}}

An important study is to analyze the effects of the {external magnetic field} on the magnetic flux. For that, we must measure such an effects in a
way that enables us to compare it to flux obtained in the absence of the {external field}. With this aim, we begin defining an effective magnetic field from the BPS equation (\ref{EQ20}) as being
\begin{equation}
B_{\text{eff}}\equiv B-g^{2}\Delta =\mp \lambda ^{2}g^{2}W\text{,}
\label{Beff}
\end{equation}
accounting for the changes in the magnetic flux due to the presence of the {external field}. We now define the effective magnetic flux inside the region $\mathcal{R}$ occupied by the compacton as,
\begin{equation}
\Phi _{\text{eff}}=\int_{\mathcal{R}}B_{\text{eff}}\,d^{2}\mathbf{x}%
=\Phi_{B}-\Phi _{\Delta }\text{,}  \label{FluxEff}
\end{equation}%
where $\Phi_B$ provides the total magnetic flux,
\begin{equation}
\Phi_{B}=\int_{\mathcal{R}}B\, d^{2}\mathbf{x}\text{,}  \label{FluxInt}
\end{equation}%
and the $\Phi_\Delta$ gives the {external field} contribution,
\begin{equation}
\Phi _{\Delta }=g^{2}\int_{\mathcal{R}}\Delta\, d^{2}\mathbf{x}\text{.}
\end{equation}
We also can directly compute the effective magnetic flux in the presence of an {external magnetic field} through the integral
\begin{equation}
\Phi _{\text{eff}}=\mp \lambda ^{2}g^{2}\int_{\mathcal{R}}W\,d^{2}\mathbf{x}%
\text{.}  \label{FluxE}
\end{equation}

{We consider it interesting} to highlight that the effective magnetic flux (\ref{FluxEff}) is associated with the existence of a magnetization \cite{MAGNETO} {produced by the presence of the external magnetic field}, and, this way, the BPS compact skyrmions described here must have a nonlinear ferromagnetic behavior. {Refs. \cite{MAGNETO, MAGNETO2} present a detailed study} of the magnetothermodynamics properties and phase transitions of the BPS-gauged baby Skyrme model under a constant external magnetic field.

\subsection{Radially symmetric skyrmions}

Let us now to study the compact skyrmions possessing a radial symmetry.
Thus, the region $\mathcal{R}$ will be a circle of radius $R$ defining the compacton's size. Then, without loss generality, we set the unitary vector as $\hat{n}=\left( 0,0,1\right) $ such that $\phi _{n}=\phi_{3}$, and we assume the hedgehog ansatz for the Skyrme field
\begin{equation}
\vec{\phi}\left( r,\theta \right) =\left(
\begin{array}{c}
\sin f(r)\cos N\theta \\[0.2cm]
\sin f(r)\sin N\theta \\[0.2cm]
\cos f(r)%
\end{array}%
\right) \text{,}  \label{EQ34}
\end{equation}%
being $r$ and $\theta $ polar coordinates, $N=\deg [\vec{\phi}]$ the winding
number introduced in (\ref{N}), and $f(r)$ a regular function satisfying the
boundary conditions
\begin{equation}
f(0)=\pi, \quad f(R)=0\text{.}  \label{EQ35}
\end{equation}%
We now redefine the $\phi _{3}$ field \cite{adam2} as%
\begin{equation}
\phi _{3}=\cos f\equiv 1-2h\text{,}
\end{equation}%
with the field $h=h(r)$ obeying%
\begin{equation}
h(0)=1, \quad h(R)=0\text{.}  \label{EQ36}
\end{equation}%
For the gauge field we consider the ansatz%
\begin{equation}
A_{i}(\mathbf{x})=-\epsilon _{ij}{x}_{j}\frac{Na}{r^{2}}\text{,}
\end{equation}%
where $a=a(r)$ is a well-behaved function that satisfies the boundary
conditions%
\begin{equation}
a(0)=0, \quad a\left( R\right)=a_{R}\text{,}  \label{EQ37}
\end{equation}%
being $a_{R}$ a finite constant.

Taking these considerations, the BPS equations are written as%
\begin{equation}
B=\frac{N}{r}\frac{da}{dr}=g^{2}\Delta \mp \lambda ^{2}g^{2}W\text{,}
\label{EQA}
\end{equation}%
\begin{equation}
Q=\frac{2N}{r}\left( 1+a\right) \frac{dh}{dr}=\mp \frac{1}{2}W_{h}\text{,}
\label{EQB}
\end{equation}%
where $W_{h}=dW/dh$ and with the {external magnetic field} being a function
depending solely on the radial coordinate, $\Delta (\mathbf{x})\equiv \Delta
(r)$. The superpotential $W\left( h\right) $ is a smooth function satisfying
the boundary conditions%
\begin{equation}
\lim_{r\rightarrow 0}W(h)=W_{0}\text{,} \quad \lim_{r\rightarrow R}W(h)=0%
\text{,\quad} \lim_{r{\rightarrow R}}W_{h}=0\text{,}
\end{equation}%
being $W_{0}$ a positive constant.

Further, the corresponding BPS potential (\ref{POT}) results%
\begin{equation}
U\left( h,\Delta \right) =\frac{\lambda ^{2}}{8}W_{h}^{2}+\frac{1}{2}%
g^{2}\left( \lambda ^{2}W\mp \Delta \right) ^{2}\text{,}  \label{Uh}
\end{equation}%
and the BPS energy density reads%
\begin{equation}
\mathcal{\varepsilon }_{_{\text{BPS}}}=\frac{\lambda ^{2}}{4}%
W_{h}^{2}+\lambda ^{4}g^{2}W^{2}\mp \lambda ^{2}g^{2}\Delta W\text{.}
\label{EQ40}
\end{equation}%
We observe here that the BPS potential and energy density associated with
the gauged BPS baby Skyrme model are recovered in the absence of the {external magnetic field} \cite{adam2}. Also, we have the Bogomol 'ny bound (\ref{EQ18B}) given by%
\begin{equation}
E_{_{\text{BPS}}}=\pm 2\pi \lambda ^{2}NW_{0}\text{,}  \label{Ebps}
\end{equation}%
where the upper (lower) sign corresponds to positive (negative) winding
number $N$ describing the skyrmions (anti-skyrmions) configurations.

For the total magnetic flux (\ref{FluxInt}) we get%
\begin{equation}
\Phi _{B}=2\pi Na_{R}\text{,}  \label{FluxFull}
\end{equation}%
where the parameter $a_{R}\in \mathds{R}$ can be computed numerically or, in
specific cases, obtained analytically {\cite{adam2}}. Further, to conduct a
detailed study as an afterthought, it is interesting to seek an expression
providing the behavior of the constant $a_{R}$. In this sense, starting from
the BPS equations (\ref{EQ20}) and (\ref{EQ21}), by using the respective
boundary conditions and after some algebraic manipulation, we obtain%
\begin{equation}
a_{R}=-1+\exp \left( -g^{2}\lambda ^{2}I_{W}\mp g^{2}I_{\Delta }\right)
\text{,}  \label{aR}
\end{equation}%
{where the parameter $I_{W}$ depending on only of the superpotential has been defined as
\begin{equation}
I_{W}=\int_{0}^{1}\frac{4W}{W_{h}}dh, \label{IWw}
\end{equation}
whereas $I_{\Delta}$ standing for the contribution coming from the {external magnetic field} is given by
\begin{equation}
I_{\Delta }=\int_{0}^{R}\frac{4\Delta }{W_{h}}\left( \frac{dh}{dr}\right) dr\text{.} \label{IDd}
\end{equation}}

In the following sections, we perform a detailed analysis of the main
features presented by the compact BPS skyrmions, including numerical
solutions in which we describe new effects and properties physical.  For this, we fix a superpotential capable of engendering compact solutions
and select some magnetic impurities.

\section{Compact skyrmions in the presence of some {external magnetic fields}\label{Sec04}}

Once our study is focused only on compact skyrmions, we consider the
following superpotential
\begin{equation}
W(h)=W_{0}h^{\gamma },\quad 1<\gamma <2, \label{EQ46}
\end{equation}%
where such a constraint to $\gamma$ values ensure the existence of the
compactons. {One points out that although we here assume the $\gamma$ values restricted by Eq. (\ref{EQ46}), such a superpotential also supports non-compact solitons for $\gamma \geq 2$, as reported in the literature.\cite{Casana045022, Casana045018, Casana065009, Casana220906309}.}

In what follows, we choose the following superpotential
\begin{equation}
W(h)=\frac{h^{3/2}}{\lambda ^{2}}\text{,}  \label{Wh32}
\end{equation}%
which induces a BPS potential (\ref{Uh}) that behaves as $U\sim h$ when $r\rightarrow R$, with $R$ defining the compacton radius. The potential vacuum behavior recalls the so-called ``old baby Skyrme potential" that also engenders compact skyrmions \cite{compacto,adam2}.

The BPS bound (\ref{Ebps}) under the considerations above leads to
\begin{equation}
E_{_{\text{BPS}}}=\pm 2\pi N\text{,}  \label{E0bps}
\end{equation}%
being therefore quantized in units of $2\pi $. Hereafter, without loss of
generality, we will take the upper sign associated with the positive winding
number $N$ describing the skyrmions configurations.

%%%%%%%%%%%%%%%%%%%%%%%%%%%%%%%%%%%%%%%%%%%%%%%%
%%%%%%%%%%%%%%%%%%%%%%%%%%%%%%%%%%%%%%%%%%%%%%%%
%%%%%%%%%%%%%%%%%%%%%%%%%%%%%%%%%%%%%%%%%%%%%%%%

\subsection{Step-type function\label{stepI}}

We follow our analysis by considering the {external magnetic field} described by the step-type function%
\begin{equation}
\Delta (r)=c\left( 1-\frac{r^{2}}{r_{0}^{2}}\right) ^{d}\theta (r_{0}-r) \text{,}
\label{EQ47}
\end{equation}
where the parameters $c\in \mathds{R}$,  $d\geq 0$, and $r_0>0$  control the {external magnetic field} features. The function $\theta (r_{0}-r)$ stands for the step function, with $r_{0}$ defining the radius {of the region occupied by the external magnetic field}.

\begin{figure*}[] \centering{\rule{0.\linewidth\includegraphics[width=8.3cm]{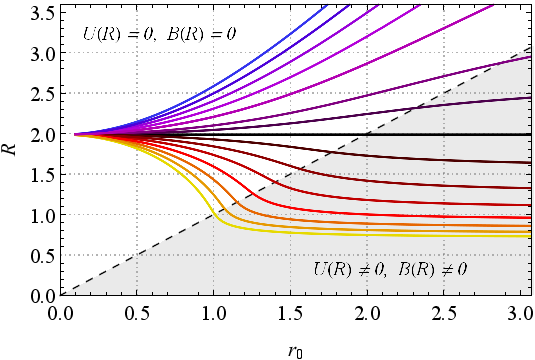}}{4cm}} {\hspace{0.1cm}}\rule{0.\linewidth\includegraphics[height=5.59cm]{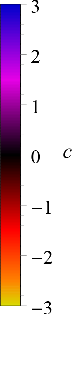}}{4cm}
	\vspace{-0.2cm}
\caption{The compacton's radius $R$ as a function of the {radius $r_{0}$} {for the {external field}}  (\ref{EQ47}) when one fixes $d=1.3$ and varies $c$ values. The dashed line delimits the solutions {possessing radii $R>r_0$ (upper region) and radii $R<r_0$} (lower region). The value $c=0$ {(horizontal black line)} stands for the BPS compacton obtained when the {external magnetic field} is absent. We also have fixed the values $N=1$, $\lambda=1$, and $g=1$.}	\label{Rxr0}
\end{figure*}

\begin{figure*}[]
\centering{\rule{0.\linewidth\includegraphics[width=8.cm]{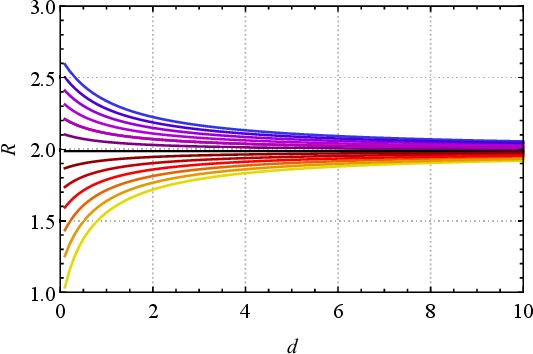}}{4cm}}
	\hspace{0.3cm}
	\centering{\rule{0.\linewidth\includegraphics[width=8.cm]{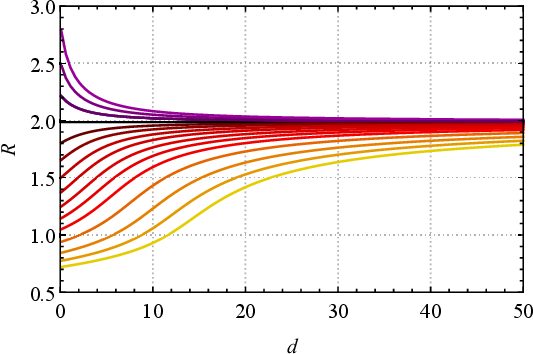}}{4cm}}
	{\hspace{0.cm}}\rule{0.\linewidth\includegraphics[height=5.31cm]{Bar.eps}}{4cm}
	\vspace{-0.2cm}
\caption{The compacton's radius $R$ as a function of the parameter $d$ with different $c$ values {for the {external field}} (\ref{EQ47}),  where we have assumed $r_{0}=0.7$ (left panel)  and $r_{0}=2.5$ (right panel).  The value $c=0$ (black line) represents the BPS compact skyrmions without {external magnetic field.} We also have fixed the values $N=1$, $\lambda=1$, and $g=1$ .}
	\label{Rxd}
\end{figure*}

{The principal} motivation for choosing the function (\ref{EQ47}) is the possibility to describe the {two scenarios} introduced in Sec. \ref{Sec03}, allowing us to explore the different effects on compact skyrmions. {Likewise, step-type functions also have been utilized} in the description of kink (or antikink) topological configurations in the presence of impurities \cite{AdamKQ} or in the modeling of two-layer ferromagnetic structures \cite{Ekomasov}.

Thereby, the {external field} (\ref{EQ47}) supports both conditions (\ref{Imp0}) and (\ref{Imp2}), i.e.,
\begin{itemize}
\item[(i)] for $r_{0}<R$, its domain satisfies the condition $\mathcal{R}_{0} \subset \mathcal{R}$, describing a magnetic impurity located totally inside the compact skyrmion (see Sec. \ref{internal}); and
\item[(ii)] for $r_{0}> R$, its domain fulfills $\mathcal{R} \subset \mathcal{R}_{0}$, depicting {an external magnetic field occupying a region larger} than that of the compacton (see Sec. \ref{extended}).
\end{itemize}

Then, by regarding the superpotential (\ref{Wh32}) and the {external field} (\ref{EQ47}), the corresponding BPS equations (\ref{EQA}) and (\ref{EQB}) reads as
\begin{equation}
\frac{N}{r}\frac{da}{dr}+g^{2}h^{3/2}-g^{2}c\left( 1-\frac{r^{2}}{r_{0}^{2}}%
\right) ^{d}\theta \left( r_{0}-r\right) =0\text{,}  \label{bps1}
\end{equation}%
\begin{equation}
N\frac{\left( 1+a\right) }{r}\frac{dh}{dr}+\frac{3}{8\lambda ^{2}}h^{1/2}=0%
\text{.}  \label{bps2}
\end{equation}%

We present below the behavior of the field profiles near the boundaries, i.e., at $r=0$ and $r=R$, and the numerical solutions attained by solving the BPS equations (\ref{bps1}) and (\ref{bps2}), under boundary conditions (\ref{EQ36}) and (\ref{EQ37}).

%%%%%%%%%%%%%%%%%%%%%%%%%%%%%%%%%%%%%%%%%%%%%%%%
%%%%%%%%%%%%%%%%%%%%%%%%%%%%%%%%%%%%%%%%%%%%%%%%

\subsubsection{The compacton radius}

{We now present how} the compacton's radius behaves {in relation to the parameters of the external magnetic field}. {This way, firstly,} Fig. \ref{Rxr0} depicts the compacton's radius {$R$ as a function of the radius $r_0$. It reveals how the {external field} (\ref{EQ47}) affects the compacton's size while keeping a fixed value for $d$ (here $=1.3$) and considering different values for $c$.} Besides, there is highlighted the boundary $r_{0}=R$ (dashed black line) that delimits the regions $r_0<R$ and $r_0> R$. The upper region contains the radii of the compact skyrmions satisfying the boundary conditions (\ref{bcBQU}), and the lower region includes the radii of the ones satisfying the conditions (\ref{bcBUext}). From the figure, we observe that for the case {when $r_0<R$,} there is a minimum value $c_{\text{min}}$ {(here $\simeq -7.631$)} such that for $c>c_{\text{min}}$ the boundary conditions (\ref{bcBQU}) are satisfied. Conversely, for the {case when $r_0> R$,} we have a maximum value $c_{\text{max}}$ (here $\simeq$ 0.316) such that for $c<c_{\text{max}}$, the boundary conditions (\ref{bcBUext}) are fulfilled. Therefore, the numerical analysis has allowed us to verify that, depending on whether {the external magnetic field is totally inside the compacton or occupying a region larger than it,} not all the $c$ values are permitted. Likewise, in Fig. \ref{Rxd}, we have presented the compacton's radius $R$ as a function of the parameter $d$ by setting {$r_0=0.7$ (left panel) and $r_0=2.5$.} We observe in both cases that, for sufficiently large $d$ values, the radius of the compact skyrmion tends to the value attained in the {absence of the external {magnetic} field} ($c=0$, black line).

%%%%%%%%%%%%%%%%%%%%%%%%%%%%%%%%%%%%%%%%%%%%%%%%
%%%%%%%%%%%%%%%%%%%%%%%%%%%%%%%%%%%%%%%%%%%%%%%%

\subsubsection{Compact skyrmions}

Firstly, for the behavior of the Skyrme profile $h(r)$ near the origin ($r\rightarrow 0$) we get
\begin{equation}
h(r)\approx1-\frac{3}{2^{4}N\lambda ^{2}}r^{2}
 +\frac{3}{2^{6}N^{2}\lambda^{2}} \left[ \frac{3}{2^{4}\lambda ^{2}}+g^{2}(c-1)\right] r^{4}. \label{h0}
\end{equation}%

\begin{figure*}[ ]
	\centering
	{\rule{0.\linewidth \includegraphics[width=8.cm]{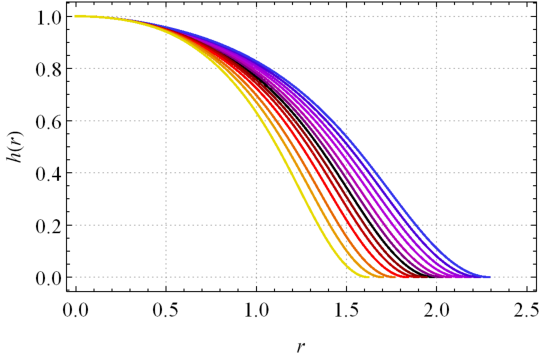}}{4cm}} {\ \hspace{0.3cm}\rule{0.\linewidth\includegraphics[width=8.cm]{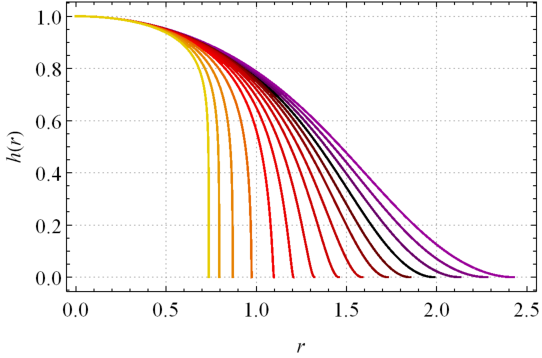}}{4cm}} {\ \hspace{-0.1cm} \rule%
		{0.\linewidth\includegraphics[width=1.04cm]{Bar.eps}}{4cm}}
	\caption{{The skyrmion} profiles $h(r)$ obtained for the {external field} (\ref{EQ47}). We have considered a fixed $d=1.3$ and different values of the parameter $c$, with $r_{0}=0.7$ (left panel) for describing the {internal case} ($r_{0}<R$) and $r_{0}=2.5$ (right panel) for representing the {extended case} ($r_{0}> R$). We also have fixed the values $N=1$, $\lambda =1$, and $g=1$.}	\label{Figh}
\end{figure*}

Already close to the compacton's border ($r\rightarrow R$), it behaves as
\begin{equation}
h(r)\approx \mathcal{H}_{R}\rho ^{2} - \frac{\mathcal{H}_{R}}{R}\rho^{3},\quad r_{0}<R,  \label{EQ60a}
\end{equation}
for the {internal case,} whereas we have for the extended one
\begin{equation}
h(r)\approx \mathcal{H}_{R}\rho ^{2}-\frac{\mathcal{H}_{R}}{R}\left[ 1-\frac{%
g^{2}R^{2}\Delta (R)}{N(1+a_{R})}\right] \rho ^{3},\quad r_{0}> R, \label{EQ60b}
\end{equation}
where $\rho =R-r\rightarrow 0^{+}$, being {$%
\mathcal{H}_{R}$ and $\Delta (R)$ given by}%
\begin{equation}
\mathcal{H}_{R}=\frac{9R^{2}}{2^{8}\lambda ^{4}N^{2}(1+a_{R})^{2}},
\label{EQ61a}
\end{equation}%
\begin{equation}
\Delta (R)\equiv\lim_{r\rightarrow R}\Delta (r)=c\left( 1-\frac{R^{2}}{%
r_{0}^{2}}\right) ^{d}\text{,}  \label{DeltaR}
\end{equation}
respectively.

Then, near the origin, Eq. (\ref{h0}) shows the {external magnetic field} has a weak influence on the $h(r)$ profile, as verified by numerical results shown in the top panels in Fig. \ref{Figh}. In turn, close to the compact's border ($r=R$), the equations (\ref{EQ60a}) and (\ref{EQ60b}), at the $\rho^3$-order, illustrate how the skyrmion's profiles approach to the vacuum value $h(R)=0$. We observe that the internal case has no explicit contribution of {$\Delta (r)$}, as shown by Eq. (\ref{EQ60a}), whereas Eq. (\ref{EQ60b}) {exhibits such a contribution for} the extended case. Specifically, in the left panel of Fig. \ref{Figh}, we have considered an {external field} with radius $r_0=0.7$ and represented the compactons with radii $R>r_0$, satisfying the conditions (\ref{bcBQU}), for the $c>c_{\text{min}}$ allowed values. On the other hand, in the right panel of Fig. \ref{Figh}, we have selected an {external field} with radius $r_0=2.5$ and shown the compact skyrmions whose radii satisfy $R< r_{0}$ fulfilling the conditions (\ref{bcBUext}), for the  $c<c_{\text{max}}$ allowed values. Furthermore, we again point out that allowed values for the parameter $c$ depend on the {external function's radius} $r_0$, which defines {the scenarios established} in Sec. \ref{Sec03}. In addition, the numerical analysis also reveals that the $c$ parameter controls the size of the compacton, i.e., its radius. The Fig. \ref{Rxr0} also shows these features related to the $c$ parameter.

%%%%%%%%%%%%%%%%%%%%%%%%%%%%%%%%%%%%%%%%%%%%%%%%
%%%%%%%%%%%%%%%%%%%%%%%%%%%%%%%%%%%%%%%%%%%%%%%%

\subsubsection{Gauge field profiles}

For the gauge field, we have the behavior close to the origin given as
\begin{equation}
a(r)\approx \frac{g^{2}(c-1)}{2N}r^{2}+\frac{g^{2}}{4N}\left( \frac{9}{%
2^{5}\lambda ^{2}N}-\frac{cd}{r_{0}^{2}}\right) r^{4}\text{,}  \label{a0}
\end{equation}%
and near the compacton's border ($r=R$), it behaves as
\begin{equation}
a(r)\approx  a_{R}+\mathcal{A}_{R}\rho ^{4}-\frac{2\mathcal{A}_{R}}{R}\rho
	^{5},\quad r_{0}<R , \label{aR1}
\end{equation}
for the {internal case, and {for the extended case}} reads
\begin{eqnarray}
a(r) &\approx & a_{R}-\frac{g^{2}R\Delta (R)}{N}\rho  \notag \\[0.2cm]
&&\hspace{-0.0cm} +\frac{g^{2}\left[ r_{0}^{2}-(2d+1)R^{2}\right]\Delta(R)}{2N \left( r_{0}^{2}-R^{2}\right) }\rho ^{2},\quad r_{0}>R ,\quad \label{aR2}
\end{eqnarray}
where $\Delta (R)$ is given in (\ref{DeltaR}) and the constant $\mathcal{A}%
_{R}$ has been defined as%
\begin{equation}
\mathcal{A}_{R}=\frac{3^{3}R^{4}g^{2}}{2^{14}\lambda ^{6}N^{4}\left(
1+a_{R}\right) ^{3}}.  \label{EQ61b}
\end{equation}

\begin{figure*}[]
	\centering
	{\rule{0.\linewidth \includegraphics[width=8.15cm]{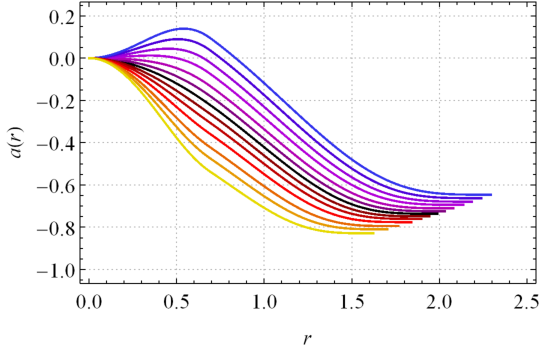}}{4cm}} {\ \hspace{0.cm%
		} \rule{0.\linewidth\includegraphics[width=8.15cm]{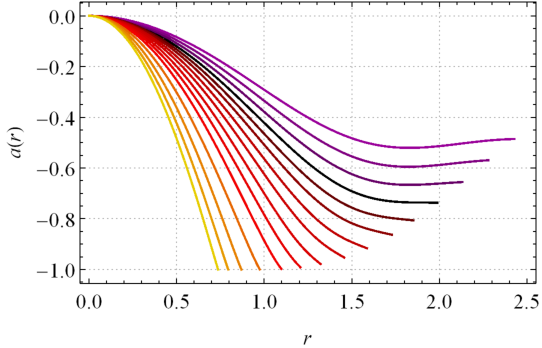}}{4cm}} {\
		\hspace{0.0cm}\rule%
		{0.\linewidth\includegraphics[width=1.032cm]{Bar.eps}}{4cm}}
	\caption{Gauge field profiles $a(r)$ obtained for the {external field} (\ref{EQ47}).  Conventions as in Fig. \ref{Figh}.}
	\label{Figa}
\end{figure*}

Around the origin, the gauge field profiles behave according to Eq. (\ref{a0}), where we observe that the {parameter $c$ affects} the way the gauge field nears the origin, whereas the behavior close to the vacuum value $a_{R}$ is given by Eqs. (\ref{aR1}) and (\ref{aR2}), for the internal and extended cases, respectively. We note that the {magnetic impurity ($r_0<R$)} relatively does not alter the behavior of the gauge profile around the {vacuum; see Eq. (\ref{aR1}).} In contrast, {in the extended scenario ($r_0>R$), such behavior acquires contributions at first order in $\rho$; see Eq. (\ref{aR2}).} Aside from this, the new effects caused by the external field are more accurately described by corresponding numerical solutions, as depicted in Fig. \ref{Figa}.

In particular, the  left panel in Fig. \ref{Figa}  shows the gauge field profile in the presence of an impurity ($r_0<R$). We observe that for a $c>0$ and sufficiently large (here $c>1$), the profile  $a(r)$ is a monotonic increasing function from $0$ up to a global maximum $a_{\text{max}}= a(r=r^{\ast})$, where  $0<r^\ast<r_0$, and then becomes a decreasing function with smooth approach to its vacuum value $a_{R}$, obeying (\ref{aR1}).  The black line represents $c=0$, the compacton gauge field profile without a magnetic impurity. For  $c<0$, the admissible values lie in the interval $c_{\text{min}}<c<0$, where the gauge profiles monotonically decrease until they attain their vacuum values. In addition, we note compacton's radius decreases as $c$ decreases towards the value $c_{\text{min}}$, with $a_{R}\rightarrow -1$ when $c\rightarrow c_{\text{min}}$. Already in the right panel of Fig. \ref{Figa}, we have depicted the compact gauge profiles in the presence of {an external field  with $r_{0}> R$ (extended case).} For $c>0$, the allowed values lie in the interval $0<c < c_{\text{max}}$ and thus, unlike the {internal case,} here the gauge profiles exhibit a global minimum when $c\rightarrow c_{\text{max}}$. Again, the value $c=0$ (black line) provides the compact gauge profile without impurities. Lastly, for $c<0$, the profile monotonically decreases towards its vacuum value {obeying (\ref{aR2})}, with $a_{R}\rightarrow -1$ when $c$ assumes sufficiently large negative values. It is noteworthy that the existence of a {maximum (internal case) or a minimum (extended case)} on the gauge field $a(r)$ profiles leads to the local flipping of the magnetic field, a fascinating feature we will discuss later.
%%%%%%%%%%%%%%%%%%%%%%%%%%%%%%%%%%%%%%%%%%%%%%%%
%%%%%%%%%%%%%%%%%%%%%%%%%%%%%%%%%%%%%%%%%%%%%%%%

\subsubsection{Magnetic field}\label{maGF}
For the total magnetic field $B(r)$ we obtain the following expression around the origin,
\begin{eqnarray}
	B(r) &\approx &g^{2}(c-1)+g^{2}\left( \frac{9}{2^{5}\lambda ^{2}N}-\frac{cd}{r_{0}^{2}}\right) r^{2}  \notag \\[0.2cm]
	&&\hspace{-0.5cm}+\left[ \frac{g^{2}cd(d-1)}{2r_{0}^{4}}-\frac{9g^{4}(c-1)}{%
		2^{7}N^{2}\lambda ^{2}}-\frac{3^{3}g^{2}}{2^{10}N^{2}\lambda ^{4}}\right]r^{4}.\quad\quad \label{B0}
\end{eqnarray}

Besides, the behavior at the boundary $r=R$  for the internal case reads as
\begin{equation}
	B(r)\approx -\mathcal{B}_{R}\rho ^{3}+\frac{3\mathcal{B}_{R}}{2R}\rho
	^{4},\quad r_{0}<R,  \label{BR1}
\end{equation}
which confirms the boundary conditions (\ref{bcBQU}), {with $\mathcal{B}_{R}$ defined by}
\begin{equation}
	\mathcal{B}_{R}=\frac{3^{3}g^{2}R^{3}}{2^{12}\lambda ^{6}N^{3}\left(
		1+a_{R}\right) ^{3}}\text{.} \label{calBr}
\end{equation}
Whereas, for the extended case, the behavior becomes
\begin{eqnarray}
	B(r) &\approx &g^{2}\Delta (R)+\frac{2g^{2}dR\,\Delta (R)}{r_{0}^{2}-R^{2}}\rho
	\notag \\[0.2cm]
	&&\hspace{-0.5cm}-\frac{g^{2}d\left[ r_{0}^{2}-(2d-1)R^{2}\right]\Delta (R) }{\left(r_{0}^{2}-R^{2}\right) ^{2}}\rho ^{2}, \quad r_{0}> R,\quad
	\label{BR2}
\end{eqnarray}%
ensuring the boundary condition (\ref{bcBUext}).

%%%%%%%%%%%%%%%%%%%%%%%%%%%%%%%%%%%%%%%%%%%%%%%
%%%%%%%%%%%%%%%%%%%%%%%%%%%%%%%%%%%%%%%%%%%%%%%
%%%%%%%%%%%%%%%%%%%%%%%%%%%%%%%%%%%%%%%%%%%%%%%

\begin{figure*}[]
\centering
{\rule{0.\linewidth \includegraphics[width=8.cm]{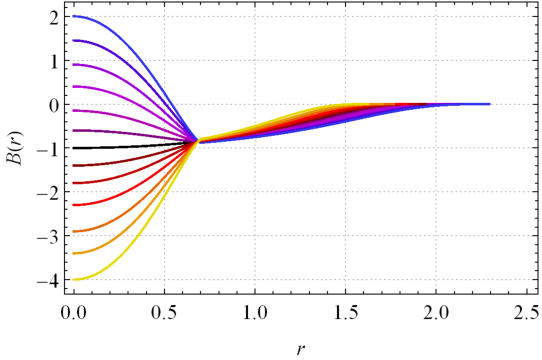}}{4cm}} {\
\hspace{0.2cm}\rule{0.\linewidth\includegraphics[width=8.cm]{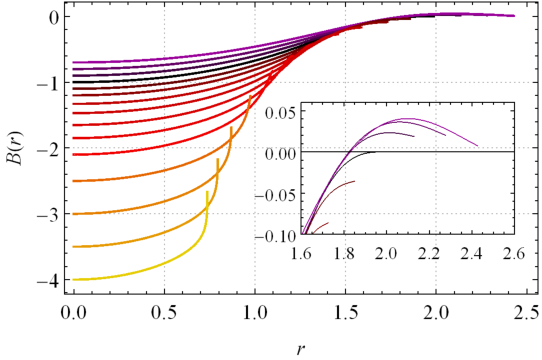}}{4cm}}
{\ \hspace{0.cm} \rule{0.\linewidth\includegraphics[width=1.045cm]{Bar.eps}}{4cm}}
\newline
{\ \hspace{-1.cm}\rule%
{0.\linewidth \includegraphics[width=8.15cm]{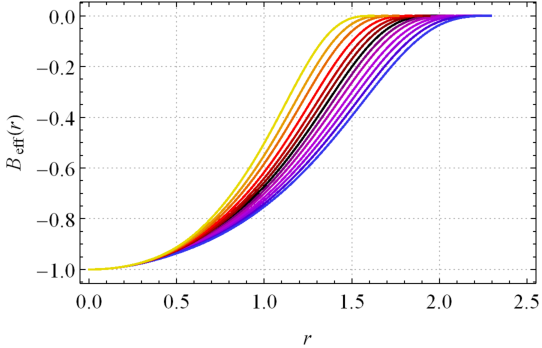}}{4cm}} {\ \hspace{%
			0.cm} \rule%
{0.\linewidth\includegraphics[width=8.15cm]{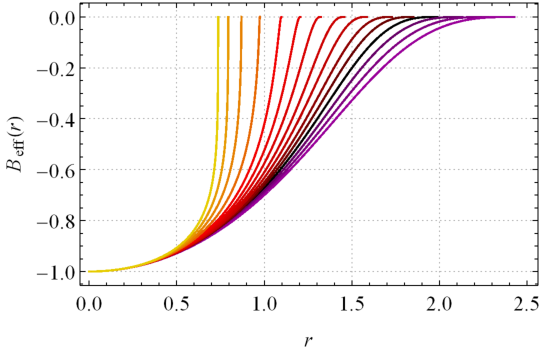}}{4cm}} {\
\hspace{0.1cm}\rule%
{0.\linewidth\includegraphics[width=1.03cm]{Bar.eps}}{4cm}}
\caption{The total magnetic field $B(r)$ (top panels) and the corresponding
effective magnetic field $B_{\text{eff}}(r)$ (bottom panels), both obtained for the {external field} (\ref{EQ47}) by considering a fixed $d=1.3$ and different values of the parameter $c$. One has assumed $r_{0}=0.7$ (left panels) for describe the {internal case} ($r_{0}<R$) and $r_{0}=2.5$ (right panels) for represent the {extended case} ($r_{0}> R$). We also have fixed the values $N=1$, $\lambda =1$, and $g=1$.}
\label{FigB}
\end{figure*}

The behavior of the total magnetic field $B(r)$ close to the frontiers is given by Eqs. (\ref{B0}), (\ref{BR1}), and (\ref{BR2}), being verified by the numerical solutions exhibited in Fig. \ref{FigB}. The left top panel shows the magnetic field profiles obtained {when consider the internal case.} Here, the numerical analysis shows a negative magnetic field for $c_{\text{min}}<c <1$,  and according to (\ref{B0}), its amplitude  at origin attains more negative values when $c\rightarrow c_{\text{min}}$. Besides, within this interval, the first derivative of the magnetic field changes its sign from positive to negative at the value
\begin{equation}
c\equiv\bar{c}_{_\text{B}}=\displaystyle\frac{9 r^2_0}{{2^{5}} d \lambda^2 N}, \label{ccbar}
\end{equation}
being here $\bar{c}_{_\text{B}}\simeq 0.106$. It happens in the radial interval $0<r<r_0$. On the other hand, for $c>1$, the magnetic field at the origin is positive and remains so along a radial region $0\leq r<r_\text{p}$ (with $r_\text{p}<r_0$) inside the impurity, besides for $r_\text{p}<r<R$ the magnetic field is negative. In addition, the first derivative of the magnetic field is always positive for $0<r<r_0$ and negative for  $r>r_0$. Therefore, the change of the magnetic field sign in the region  $0 \leq r<r_{\text{p}}$ implies a local magnetic flux flipping produced by the presence of the impurity. This phenomenon does not occur with the magnetic field of the BPS skyrmions engendered in the absence of the external magnetic field. Note that this peculiar feature is a consequence of the behavior of the gauge field's  $a(r)$ discussed previously, where emerges a global maximum for $c>1$ values. Furthermore, one observes a behavior common to the profiles, which we can describe in the following way: they start from the origin, {then converge quickly when approaching the impurity's radius $r_{0}$} and so proceed monotonously to the vacuum value at the compacton border, in agreement with (\ref{BR1}) and therefore satisfying (\ref{bcBQU}).

In the right top panel of Fig. \ref{FigB} are depicted the total magnetic field's profiles engendered in the {extended case when fixing $r_0=2.5$.} Thus, we have a magnetic field whose value at the origin is always negative due to the restriction $c<c_{\text{max}}\simeq 0.316<1$, in agreement with (\ref{B0}), that is corroborated by calculating the value (\ref{ccbar}) that results in $\bar{c}_{_\text{B}}\simeq 1.352$, a value outside the range of allowed values of $c$. Besides, the amplitude at the origin grows negatively for more negative values of $c$, whereas the compacton radius diminishes continuously. Further, as previously mentioned by discussing gauge field's profiles $a(r)$, we have the emergence of the global minimum when $c\rightarrow c_{\text{max}}$ engendering also a flipping of the magnetic field's sign (see the {amplified} depiction inside in the right top panel of Fig. \ref{FigB}). Note that the profiles obey the behavior given in (\ref{BR2}), which guarantees the boundary condition established by Eq. (\ref{bcBUext}).

\begin{figure*}[]
	\centering
	{\rule{0.\linewidth \includegraphics[width=8.05cm]{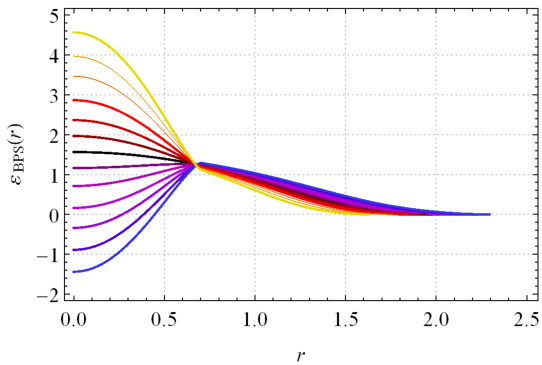}}{4cm}} {\
		\hspace{			0.2cm}\rule%
		{0.\linewidth\includegraphics[width=7.9cm]{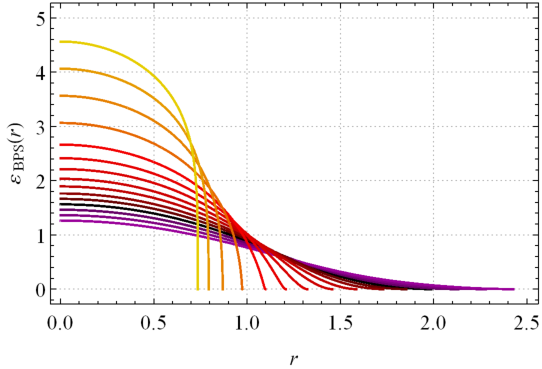}}{4cm}} {\ \ \hspace{%
			0.cm} \rule{0.\linewidth\includegraphics[width=1.055cm]{Bar.eps}}{4cm}}
\caption{The profiles of the BPS energy density $\varepsilon_{_{\text{BPS}}}(r)$ obtained for the {external field} (\ref{EQ47}). Conventions as in Fig. \ref{Figh}.}
	\label{FigE}
\end{figure*}

On the other hand, the effective magnetic field $B_\text{eff}(r)$ defined in Eq. (\ref{Beff}) {behaves when $r\rightarrow 0$ as}
\begin{eqnarray}
B_{\text{eff}}(r)&\approx& -g^{2}+\frac{9 g^{2}}{2^{5}\lambda^{2}N}r^{2} \nonumber \\[0.2cm]
&& - \frac{9g^{2}}{2^{7}N^{2}\lambda ^{2}} \left[ g^{2}(c-1)+\frac{3}{8\lambda^{2}} \right] r^{4}.\label{Beff0}
\end{eqnarray}
In contrast, {when $r\rightarrow R$, the effective magnetic field for the internal case reads}
\begin{equation}
B_{\text{eff}}(r)\approx -\mathcal{B}_{R}\rho^{3} +\frac{3\mathcal{B}_{R}}{2R}\rho^{4},\quad r_{0}<R,  \label{BR1x}
\end{equation}
{which has the same behavior as the total magnetic field, which happens because of the definition of the impurity introduced in Eq. (\ref{Imp0}). While for the extended case, the effective magnetic field approaches its vacuum value as}
\begin{eqnarray}
B_{\text{eff}}(r)&\approx&-\mathcal{B}_{R}\rho ^{3} \nonumber \\[0.2cm] &&+\frac{3\mathcal{B}_{R}}{2R} \left[ 1-\frac{R^{2}g^{2}\Delta (R)}{N\left( 1+a_{R}\right) }\right]  \rho^{4},\quad r_{0}> R,\quad \label{Beffext}
\end{eqnarray}
{where $\mathcal{B}_{R}$ is given in Eq. (\ref{calBr}). Thus, we conclude that the effective magnetic field possesses a null vacuum value, similar to what happens in the absence of an external magnetic field.}

The effective magnetic field's profiles are shown in the bottom panels in Fig. \ref{FigB}. The profiles are well-behaved around the origin for {both scenarios,} in agreement with (\ref{Beff0}). In general, the profile format follows a similar behavior to the one emerging in the absence {of the external field.} Nevertheless, the {external field} affects the size of the compactons, i.e., their radii, as shown by the bottom panels in Fig. \ref{FigB}. Specifically, the radius of the compacton decreases as $c$ grows negatively {in both the internal and extended cases,} such as previously observed through Fig. \ref{Rxr0}. The diminishing of the compacton's radius can be related to the reduction in the area occupied by the magnetic field, thus impacting the total magnetic flux carried by the {compact skyrmion}.

%%%%%%%%%%%%%%%%%%%%%%%%%%%%%%%%%%%%%%%%%%%%%%%%
%%%%%%%%%%%%%%%%%%%%%%%%%%%%%%%%%%%%%%%%%%%%%%%%

\subsubsection{BPS energy density}

The corresponding BPS energy density (\ref{EQ40}), for the superpotential (\ref{Wh32}) and {external field} (\ref{EQ47}), behaves near the origin as
\begin{eqnarray}
\varepsilon _{_{\text{BPS}}} &\approx &\frac{9}{{2^{4}}\lambda ^{2}}-g^{2}(c-1)
\notag \\[0.2cm]
&& -\left[ \frac{3^{3}}{2^{8}N\lambda ^{4}}-\frac{9g^{2}(c-2)}{2^{5} N\lambda^{2}}-\frac{cdg^{2}}{r_{0}^{2}}\right] r^{2}\text{.}\quad  \label{En0}
\end{eqnarray}
Moreover, at the compacton's border, {for the internal case,} it reads as
\begin{equation}
\varepsilon _{_{\text{BPS}}}\approx \mathcal{E}_{R}\rho ^{2}-\frac{\mathcal{E%
}_{R}}{R}\rho ^{3},\quad r_{0}<R,
\end{equation}%
and, {for the extended {case}, the BPS} energy density behaves as
\begin{equation}
\varepsilon _{_{\text{BPS}}}\approx \mathcal{E}_{R}\rho ^{2}-\frac{\mathcal{E%
}_{R}}{R}\left[ 1-\frac{2R^{2}g^{2}\Delta (R)}{3N\left( 1+a_{R}\right) }%
\right] \rho ^{3},\quad r_{0}> R\text{,} \label{EbpsR}
\end{equation}
where we have defined $\mathcal{E}_{R}$ by
\begin{equation}
\mathcal{E}_{R}=\frac{3^{4}R^{2}}{2^{12}\lambda ^{6}N^{2}\left(
1+a_{R}\right) ^{2}}.
\end{equation}
{We thus verify that the BPS energy density has a null vacuum value, satisfying the condition proposed in Eq. (\ref{BCEn}), which is also observed in the absence of the external field.}

\begin{figure*}[]
\centering
{\rule{0.\linewidth \includegraphics[width=7.9cm]{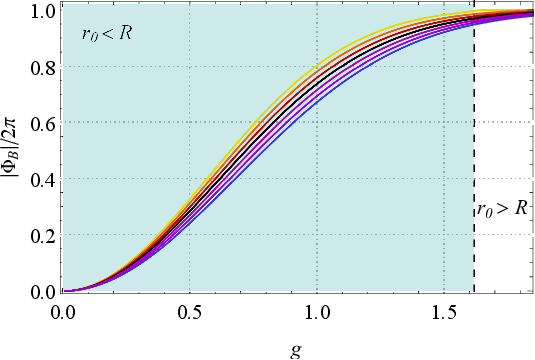}}{4cm}} {\
	\hspace{		0.4cm}\rule%
	{0.\linewidth\includegraphics[width=7.9cm]{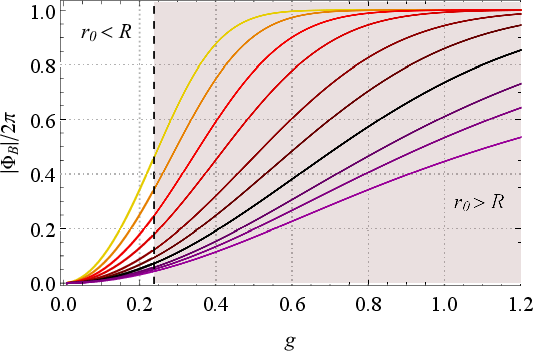}}{4cm}} {\ \ \hspace{%
		-0.1cm} \rule{0.\linewidth\includegraphics[width=1.03cm]{Bar.eps}}{4cm}}%
\newline
{\ \vspace{0.cm}\rule%
	{0.\linewidth \includegraphics[width=7.9cm]{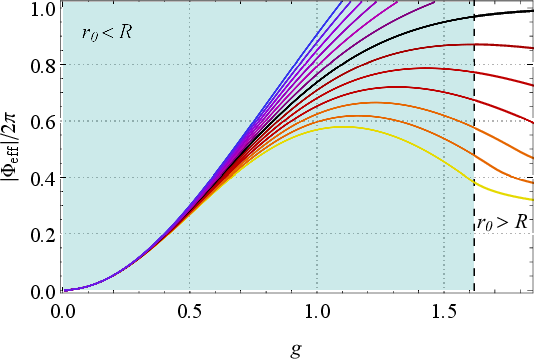}}{4cm}} {\ \hspace{%
		0.2cm	} \rule{0.\linewidth\includegraphics[width=8.cm]{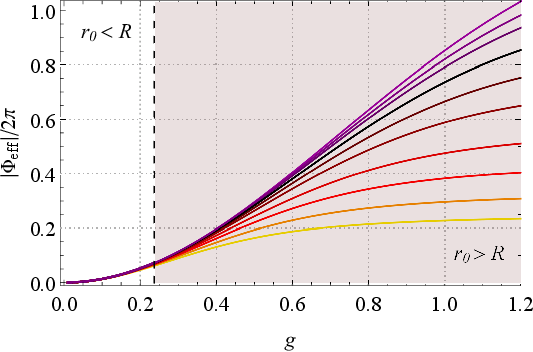}}{4cm}} {%
\ \hspace{0.1cm}\rule
{0.\linewidth\includegraphics[width=1.045cm]{Bar.eps}}{4cm}}
\caption{The total flux magnetic $\Phi _{B}$ (top panels) and the corresponding effective magnetic flux $\Phi _{\text{eff}}$ (bottom panels) are both displayed as functions of the electromagnetic coupling $g$ for the  {external field} (\ref{EQ47}) when one fixes $d=1.3$ and runs the $c$ values. {Besides,  for the internal {case} was fixed $r_{0}=0.6$ (left panels), and for the extended case $r_{0}=3.5$ (right panels).} We also have fixed the values $N=1$ and $\lambda=1$.}
\label{FvsG}
\end{figure*}

Figure \ref{FigE} depicts the numerical profiles of the BPS energy density (\ref{EQ40}), so the left panel shows the ones obtained in the presence of {a magnetic impurity (internal scenario),} {and the right panel depicts the profiles emerging for the extended scenario.} One can better understand the profiles' format by looking at the behavior given in Eq. (\ref{En0}) and by its first derivative. Firstly, we see that the amplitude at the origin becomes null when $c$ attains the value
\begin{equation}
c\equiv\bar{c}_{\varepsilon}=1+\frac{9}{{2^{4}} \lambda^2 g^2} , \label{cbarE}
\end{equation}
independent of {the parameters $d$ and $r_0$.} Besides, the first derivative of (\ref{En0}) at the origin changes its sign, i.e., the second derivative becomes null em $r=0$ when $c$ assumes the value
\begin{equation}
\bar{\bar{c}}_{\varepsilon}=\frac{9 \left({2^{4}} g^{2} \lambda^{2}+3\right) r_0^{2}}{8 \lambda^{2} g^{2} \left({2^{5}} N d \,\lambda^{2}+9 r_0^{2}\right)}, \label{cbarE1}
\end{equation}
which now depends on the {parameters $d$ and $r_0$.} This way, for the parameter values chosen to perform the numerical analysis, the value (\ref{cbarE}) is $\bar{c}_{\varepsilon}  \simeq  1.563$ that belongs to the range $c>c_{\text{min}}\simeq -7.631$ of the values allowed {in the internal case} ($r_0=0.7$). It explains the change in sign of the amplitude of the BPS energy density at the origin when $c$ grows (see the left panel in Fig. \ref{FigE}). On the other hand, in the {extended {case}} ($r_0=2.5$), the value $\bar{c}_{\varepsilon} \simeq 1.563$ is outside the interval $c<c_{\text{max}} \simeq 0.316$ of the permitted values, therefore, in this case, the amplitude at the origin never be null, such as shown by the right panel of the Fig. \ref{FigE}.  Likewise, in the {internal case} ($r_0=0.7$), the concavity, or the convexity of the profiles of the BPS energy density, in the interval $0\leq r<r_0$, is explained by the value (\ref{cbarE1}), which here is $\bar{\bar{c}}_{\varepsilon} \simeq 0.228>c_{\text{min}}\simeq -7.631$, subsequently for $c_{\text{min}}<c<0.228$ the profiles are concave and for $c>0.228$ they are convex. In contrast, in the {extended scenario} ($r_0=2.5$) the value (\ref{cbarE1}) becomes $\bar{\bar{c}}_{\varepsilon} \simeq 1.365$ resulting be greater than the maximum allowed value $c_{\text{max}} \simeq 0.316$, consequently, the profiles are concave for all permitted values $c<c_{\text{max}}$, such as observed in the right panel of Fig. \ref{FigE}.

%%%%%%%%%%%%%%%%%%%%%%%%%%%%%%%%%%%%%%%%%%%%%%%%
%%%%%%%%%%%%%%%%%%%%%%%%%%%%%%%%%%%%%%%%%%%%%%%%

\vspace{-0.2cm}

\subsubsection{Magnetic flux}

The total magnetic flux $\Phi _{B}$, given in (\ref{FluxFull}), is proportional to $a_R$ defined in Eq. (\ref{aR}), which becomes
\begin{equation}
a_{R}=-1+\exp \left( -\frac{4}{3}g^{2}\lambda ^{2}-g^{2}I_{\Delta }\right)
\text{,}  \label{aR0}
\end{equation}%
and by regarding the superpotential (\ref{Wh32}), we have calculated (\ref{IWw}) to obtain $I_{W}=4/3$. From (\ref{IDd}), the quantity $I_{\Delta}$  is given by
\begin{equation}
I_{\Delta }=c\frac{8\lambda^2}{3}\int_{0}^{R}\left( 1-\frac{r^{2}}{r_{0}^{2}}\right)
^{d} \frac{\theta(r_0-r)}{\sqrt{h}} \frac{dh}{dr}\;dr\text{,}  \label{IRd}
\end{equation}%
will be calculated numerically to perform the magnetic flux analysis.

For our purpose, in the top panels of Fig. \ref{FvsG}, we have depicted the absolute value of the total magnetic flux $\Phi_{B}$ given by (\ref{FluxFull}) as a function of $g$ for different $c$ values. With a loss of generality, we present the total flux produced in {the internal scenario} (with $r_{0}=0.6$, on the left panel) and {in the extended one} (with $r_{0}=3.5$, on the right panel).

{For sufficiently large $g$ values, in the internal case, we observe that when the parameter $c\rightarrow \bar{c}_{\text{min}}= -3.0$, the gauge field vacuum value approaches $a_{R}\rightarrow-1$. Already in the extended scenario, the parameter $c< \bar{c}_{\text{max}} =0.3$, and when it becomes more negative, the gauge field vacuum value nears $a_{R}\rightarrow-1$, too. Thus, in both situations, $\Phi_{B}\rightarrow-2\pi$, which means that in such limits, the magnetic flux becomes quantized in units of $2\pi$.}

{Additionally, in the left panels, we have highlighted the interval $0<g<g_{\text{max}} \simeq 1.619$ that, for the internal case, defines the allowed values engendering compact skyrmions whose radii satisfy the condition $R>r_0$, thus ensuring that the boundary conditions established in Eqs. (\ref{BCEn}), (\ref{Qint}), and (\ref{bcBQU}) be satisfied. Similarly, in the extended scenario (right panels), we identified the range $g>g_{\text{min}} \simeq 0.238$ containing the permitted values that generate compactons whose radii fulfill the condition $R< r_0$, hence guaranteeing the boundary conditions (\ref{BCEn}), (\ref{Qext}), and (\ref{bcBUext}) are satisfied too.} Moreover, we have also depicted the effective magnetic flux (\ref{FluxE}) for {both scenarios in} the bottom panels of Fig. \ref{FvsG}.

\begin{table}[t]
	\caption{Values for the electromagnetic coupling constant in the presence of {the external magnetic field}.}
	\begin{ruledtabular}
		\begin{tabular}{cc|cc}
			\multicolumn{2}{c}{\hspace{-0.25cm}{Internal scenario}$^a$} & \multicolumn{2}{c}{\hspace{-0.5cm}{Extended scenario}$^b$} \\ \hline
			$r_{0}$ & $g_{\text{max}}$ & {$r_{0}$} & $g_{\text{min}}$ \\
			\colrule
			&  & &  \\ [-0.25cm]	
			{1.0}&$1.005$&$3.5$&{0.238} \\
			{0.7}&$1.403$&$3.0$&{0.635} \\
			{0.6}&$1.619$&$2.5$&{0.989} \\
			{0.4}&$2.370$&$2.0$&{1.282} \\
			{0.3}&$3.133$&$1.5$&{1.545} \\
		\end{tabular}
	\end{ruledtabular}
	\par
	\vspace{-0.3cm}
	\par
	\begin{flushleft}
		{\footnotesize {We have fixed $^a$$\bar{c}_{\text{min}}=-3.0$ for {the internal case}  and $^b$$\bar{c}_{\text{max}}=0.3$ for {the extended case,} by setting $N=1$, $\lambda=1$, and $d=1.3$.} }
	\end{flushleft}\label{T1}\vspace{-0.5cm}
\end{table}

%%%%%%%%%%%%%%%%%%%%%%%%%%%%%%%%%%%%%%%%%%%%%%%%
%%%%%%%%%%%%%%%%%%%%%%%%%%%%%%%%%%%%%%%%%%%%%%%%

\subsubsection{Effects of the {external magnetic field} on  the electromagnetic coupling constant}

In the gauged BPS baby Skyrme model \cite{adam2}, the electromagnetic coupling constant $g$ controls the compacton's size. A similar situation {emerges in our model, where we perceive that, in addition to the coupling constant $g$ controlling the compacton's size, the parameters defining the external field also do so. Thus, one discerns a relationship between the allowed values of the coupling constant $g$ and the external field parameters, which are constrained or restricted by the boundary conditions established in the internal or the extended scenarios.

The table \ref{T1} helps us understand this relationship better. We observe that the electromagnetic coupling constant has restricted values, depending on whether the external function, characterized by the radius $r_0$, describes the internal or extended scenario. In this sense, for convenience, to analyze the internal case, we have fixed the value $\bar{c}_{\text{min}}$ such that the values $c>\bar{c}_{\text{min}}$ are permitted. On the other hand, for the extended case, we have fixed $\bar{c}_{\text{max}}$, so the values $c<\bar{c}_{\text{max}}$ are the ones allowed. To perform the} numerical analysis, we also set $N=1$, $\lambda=1$, and $d=1.3$. {For the internal scenario, once chosen the radius $r_0$,} there exists a maximum value $g_{\text{max}}$ of the electromagnetic coupling constant such that for $g<g_{\text{max}}$, the condition $r_0<R$ remains valid and the compacton achieves the boundary conditions established in Eqs. (\ref{BCEn}), (\ref{Qint}), and (\ref{bcBQU}). Similarly, {in the extended scenario, once a radius $r_0$ is selected,} the electromagnetic coupling constant attains a minimum value $g_{\text{min}}$, {then the condition $r_0> R$ is only fulfilled when $g>g_{\text{min}}$, and, therefore, the compacton satisfies the boundary conditions specified} in Eqs. (\ref{BCEn}), (\ref{Qext}), and (\ref{bcBUext}).

\begin{figure*}[]
	\centering
	{\rule{0.\linewidth \includegraphics[width=7.8cm]{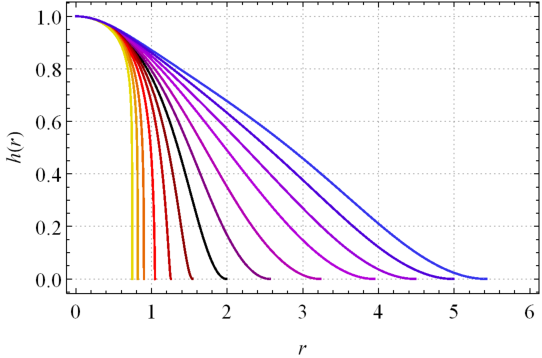}}{4cm}} {\
		\hspace{0.2cm}\rule%
		{0.\linewidth\includegraphics[width=8.cm]{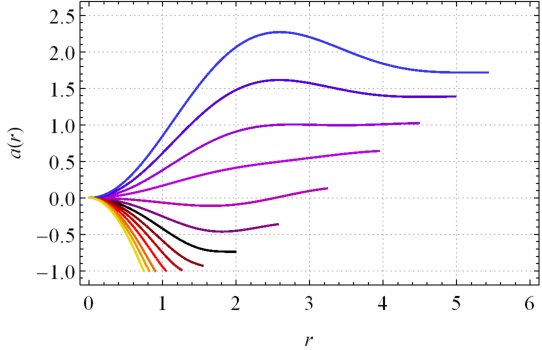}}{4cm}} {\ \hspace{0.cm} \rule{0.\linewidth\includegraphics[width=1.02cm]{Bar.eps}}{4cm}}
	\caption{The skyrmion profiles $h(r)$ (left panel) and the gauge field profiles $a(r)$ (right panel), both obtained for the {external field}  (\ref{ImP2}) when one sets $\sigma=1.3$ and runs the values of the parameter $c$. We also have fixed the values $N=1$, $\lambda =1$, and $g=1$.}
	\label{FigGauss1}
\end{figure*}

\begin{figure*}[]
	\centering
	{\rule{0.\linewidth \includegraphics[width=8.cm]{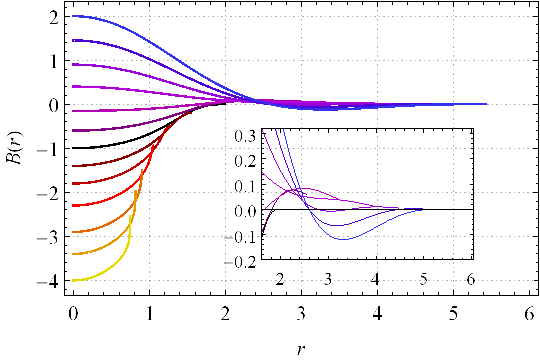}}{4cm}} {\ \hspace{%
			0.cm} \rule%
		{0.\linewidth\includegraphics[width=8.15cm]{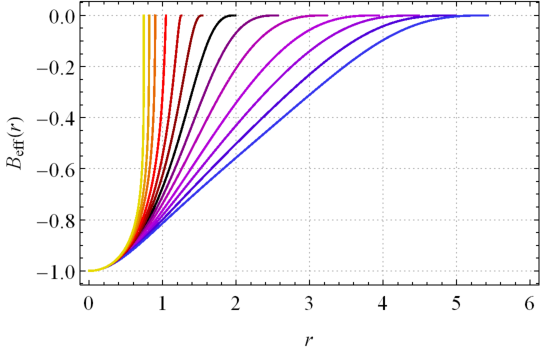}}{4cm}} {\ \hspace{%
			0.1cm}\rule{0.\linewidth\includegraphics[width=1.04cm]{Bar.eps}}{4cm}}\
	\hspace{1.cm}
	\caption{The total magnetic field $B(r)$ (left panel) and the corresponding effective magnetic field $B_{\text{eff}}(r)$ (right panel), both obtained for the {external field}  (\ref{ImP2}). Conventions as in Fig. \ref{FigGauss1}.}
	\label{FigGauss2}
\end{figure*}

\begin{figure}[b]
\centering\includegraphics[width=7.5cm]{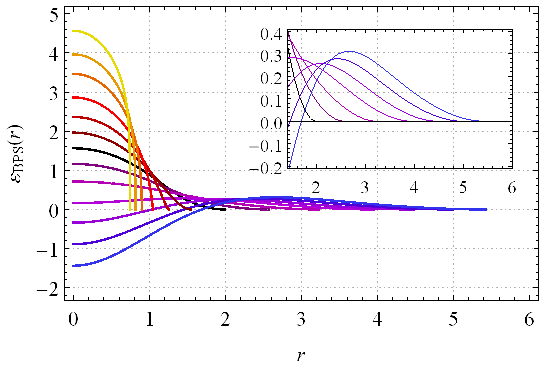} \hspace{0.05cm}\includegraphics[width=0.985cm]{Bar.eps}
	\caption{The profiles of the BPS energy density $\varepsilon_{_{\text{BPS}}}(r)$ obtained for the {external field}  (\ref{ImP2}). Conventions as in Fig. \ref{FigGauss1}.}
	\label{fig-9}
\end{figure}

{Fig. \ref{FvsG} illustrates an insightful example of how the external field's parameters restrict the allowed $g$ values. In this sense, for the internal case (left panel), we have adopted $r_0=0.6$, which fixes the maximum value $g_{\text{max}} \simeq 1.619$ for the electromagnetic coupling (see table \ref{T1}), represented by the region $r_0 <R$. Regarding the extended case (right panel), we have chosen $r_{0}=3.5$, achieving the minimum value $g_{\text{min}} \simeq 0.238$ for the electromagnetic coupling (see table \ref{T1}), represented by the region $r_{0}> R$.}

%%%%%%%%%%%%%%%%%%%%%%%%%%%%%%%%%%
%%%%%%%%%%%%%%%%%%%%%%%%%%%%%%%%%%
%%%%%%%%%%%%%%%%%%%%%%%%%%%%%%%%%%

\subsection{{Gaussian function}\label{stepII}}

{We now continue our analysis by choosing an external magnetic field defined by the Gaussian function,}
\begin{equation}
\Delta^{(g)} (r)=c\exp \left( -\frac{r^{2}}{2\sigma ^{2}}\right) ,  \label{ImP2}
\end{equation}
where $c \in \mathds{R}$ and $\sigma>0$ are constants that control the amplitude and the effective width, respectively. {The effects of the external function (\ref{ImP2}) on the formation of compact skyrmions will be analyzed by using} the approach described in Sec. \ref{extended}. {Additionally, regarding the investigation} of topological defects in the presence of magnetic impurities, the Gaussian function has been used previously, as referenced in Refs. \cite{daHora, JASK}.

{The left panel of Fig. \ref{FigGauss1} shows the profiles $h(r)$ of the compact skyrmions, which} behave around the origin according to Eq. (\ref{h0}), {while close to} the compacton's boundary, i.e., when $\rho=R-r\rightarrow 0^+$, {they comport} as given by Eq. (\ref{EQ60b}), where $\Delta (R)$ now reads as
\begin{equation}
\Delta (R)=c\exp \left( -\frac{R^{2}}{2\sigma^{2}}\right) =\lim_{r\rightarrow
R}\Delta^{(g)} (r). \label{DeltaGs}
\end{equation}
The numerical analysis shows that for $c\geq 0$, the profiles {smoothly approach the vacuum value, such that} the compacton's radius is longer than {that obtained in the absence of an external magnetic field.} Conversely, for $c<0$, they achieve the vacuum value quickly, and simultaneously, the compacton's radius becomes shorter than the one {attained without the external field.}

Similarly, the right panel of Fig. \ref{FigGauss1} depicts the gauge field profiles $a(r)$. {The calculation of the profiles' behavior near the origin provides}
\begin{equation}
a(r)\approx \frac{g^{2}(c-1)}{2N}r^{2}+\frac{g^{2}}{4N}\left( \frac{9}{%
2^{5}\lambda ^{2}N}-\frac{c}{2\sigma^{2}}\right) r^{4},
\end{equation}%
which clarifies why the profiles exhibit a convex format for $c>1$ and a concave shape for $c<1$.  {Additionally, the profiles' behavior close to} the compacton's border is given by
\begin{equation}
a(r)\approx a_{R}-\frac{g^{2}R\Delta (R)}{N}\rho -\frac{g^{2}(R^{2}-\sigma^{2})
\Delta (R)}{2N\sigma^{2}}\rho ^{2}.
\end{equation}
The numerical solutions also show that {the external field} noticeably affects the $a(r)$ profiles by promoting the emergence of a global maximum for $c>0$ and a global minimum for $c<0$.

\begin{figure*}[]
	\centering
	{\rule{0.\linewidth \includegraphics[width=8.cm]{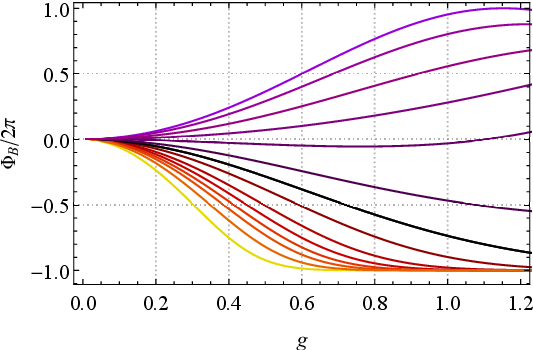}}{4cm}} {\
		\hspace{0.2cm}\rule%
		{0.\linewidth\includegraphics[width=7.8cm]{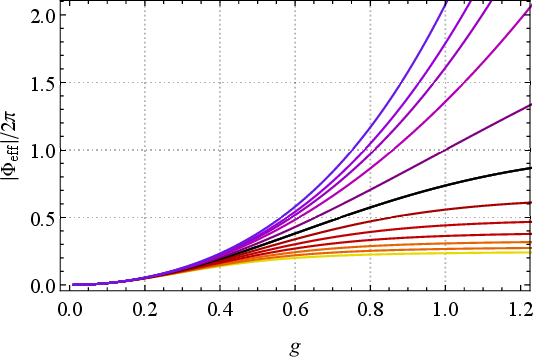}}{4cm}} {\ \
		\hspace{0.cm} \rule%
		{0.\linewidth\includegraphics[width=1.035cm]{Bar.eps}}{4cm}}
\caption{The total flux magnetic $\Phi _{B}$ (left panel) and the corresponding effective magnetic flux $\Phi _{\text{eff}}$ (right panel) are both represented as functions of the electromagnetic coupling $g$ for the {external field} (\ref{ImP2})  when one set $\sigma=1.3$ and runs the $c$ values. We also have fixed the values $N=1$ and $\lambda=1$.}
	\label{fig-10}
\end{figure*}

Figure \ref{FigGauss2} presents the magnetic field profiles (left panel). We observe that the amplitude at $r=0$ has negative values for $c<1$ and turns positive when $c>1$, as can be verified {by examining the behavior} around the origin, {which is given by}
\begin{eqnarray}
B(r) &\approx &g^{2}(c-1)+g^{2}\left( \frac{9}{32\lambda ^{2}N}-\frac{c}{%
2\sigma ^{2}}\right) r^{2}  \notag \\[0.2cm]
&&\hspace{-0cm}+\left[ \frac{g^{2}c}{8\sigma ^{4}}-\frac{9g^{4}(c-1)}{%
2^{7}N^{2}\lambda ^{2}}-\frac{3^{3}g^{2}}{2^{10}N^{2}\lambda ^{4}}\right]
r^{4}.\quad  \label{EQ55}
\end{eqnarray}
It also allows us to {explain why for} the $c<9\sigma^2/(16 N\lambda^2)$ (here $\sim 0.951$), the magnetic field around the origin acquires a convex format, {whereas that} for $c>9\sigma^2/(16 N\lambda^2)$, it presents a concave layout. Besides, for $c<1$, the magnetic field profiles present a format like the ones engendered {in the extended scenario} studied in section \ref{maGF} (see the top right panel of Fig. \ref{FigB}), with the tails also exhibiting the flipping of the magnetic field before achieving the vacuum value. Already, for $c>1$, the profiles possess a layout {similar to the} ones observed in the internal case described in the section \ref{maGF} (see the top left panel of Fig. \ref{FigB}), with the tails now showing the inversion of the magnetic field before reaching the vacuum value. The flipping of the magnetic field tail {(see the amplified depiction inside in the left panel of Fig. \ref{FigGauss2})} is observed through the profile behavior when $r\rightarrow R$, reading as
\begin{equation}
B(r)\approx g^{2}\Delta (R)+\frac{g^{2}R\Delta (R)}{\sigma^{2}}\rho +\frac{%
g^{2}(R^{2}-\sigma^{2})\Delta (R)}{2\sigma^{4}}\rho ^{2},\quad \label{BRG}
\end{equation}
{with $\Delta (R)$ defined in Eq. (\ref{DeltaGs})}, which also confirms the boundary condition (\ref{bcBUext}).

The right panel of Fig. \ref{FigGauss2} shows the effective magnetic field. The profiles behave at $r=0$ as given by Eq. (\ref{Beff0}), {which verifies that} the magnetic field amplitude is constant and will always have a convex format around {the origin's neighborhood, because the external field does not significantly affect the profiles, i.e., the behavior is similar} to that obtained without the external field. {The principal effect arising due to the presence of the external field} is changes in the length of the compacton's {radius, which becomes minor} whenever $c$ acquires more negative values and turns bigger for increasing positive values of $c$. Moreover, the tail behaves {similarly to} the one obtained {in the absence of the external field,} being described by Eq. (\ref{Beffext}). {This way, the tail of the effective magnetic field does not suffer a flip.}

Figure \ref{fig-9} displays the profiles of the BPS energy {density, whose respective behaviors at $r=0$ read as}
\begin{eqnarray}
\varepsilon _{_{\text{BPS}}} &\approx &\frac{9}{{2^{4}}\lambda ^{2}}-g^{2}(c-1)
\notag \\[0.2cm]
&&-\left[ \frac{27}{{2^{8}}N\lambda ^{4}}-\frac{9g^{2}(c-2)}{{2^{5}}N\lambda ^{2}}-\frac{cg^{2}}{2\sigma ^{2}}\right] r^{2}. \label{EnG}
\end{eqnarray}
It illuminates why the BPS energy density's amplitude at $r=0$ changes its sign when $c\neq \bar{c}_ {\varepsilon}$ (here $\bar{c}_{\varepsilon}\simeq 1.563$) in according to Eq. (\ref{cbarE}). Similarly, the second derivative is null when $c$ assumes the value
\begin{equation}
c_{_\text{G}} =\frac{9({2^{4}}g^{2}\lambda ^{2}+3)\sigma ^{2}}{{2^{3}}g^{2}\lambda ^{2}\left(9\sigma ^{2}+{2^{4}}N\lambda ^{2}\right) }.
\end{equation}
This way, around the origin's, the profile possesses a convex format for $c>c_{_\text{G}}$ (here $c_{_\text{G}}\sim 1.157$) and gains a concave layout for $c<c_{_\text{G}}$. Furthermore, for $\rho=R-r \rightarrow 0^+$, the BPS energy density obeys the same expression in Eq. (\ref{EbpsR}).
{We observe that for $c>\bar{c}_{\varepsilon}$, despite the BPS energy density near the origin having a negative value, the profiles suffer a sign inversion {(see the amplified depiction inside of Fig. \ref{fig-9})} such that the tails always approach the vacuum values, maintaining a positive sign.}

Finally, we have presented in Fig. \ref{fig-10} the profiles of the total (left panel) {and effective} (right panel) magnetic flux, respectively, {for a fixed} $\sigma$, as functions of the electromagnetic coupling $g$ and the $c$ parameter. The total magnetic flux $\Phi _{B}$ is negative for positive small values of $c$; in contrast, for sufficiently large positive $c$, it becomes positive such that for increasing values of $c$ and $g$, the gauge vacuum value, $a_{R} \rightarrow 1$ then $\Phi_{B}\rightarrow 2\pi$. On the other hand, for $c<0$, the magnetic flux is always negative, and for sufficiently large negative $c$ values and large enough values of $g$, the gauge field vacuum value $a_{R}\rightarrow -1$, thus $\Phi_{B} \rightarrow -2\pi$. Such a situation contrasts totally {with results attained with a finite-sized external function} (\ref{EQ47}) [see the top panels in Fig. \ref{FvsG}], where the vacuum value $a_{R}$ always goes to $-1$ such that $\Phi_{B} \rightarrow -2\pi$.   On the other hand, the effective magnetic flux $\Phi_{\text{eff}}$  displays a behavior very similar to the one presented by {the extended scenario built starting from the external function} (\ref{EQ47}) [see the right {bottom panel} in Fig. \ref{FvsG}].

%%%%%%%%%%%%%%%%%%%%%%%%%%%%%%%%%%
%%%%%%%%%%%%%%%%%%%%%%%%%%%%%%%%%%

\section{Remarks and conclusions\label{Sec05}}

We have investigated the emergence of BPS compactons {in a gauged} restricted baby Skyrme model interacting with {an external magnetic field. For this purpose,} we have focused on constructing the BPS structure, {which enables us} to achieve the BPS equations whose solutions saturate the Bogomol'nyi bound and set the self-dual potential. {While trying to implement the BPS formalism, it becomes apparent that its} construction depends on the relative size between the compacton and {  spatial extent of the external magnetic field}. To perform such an analysis, {  we have investigated} two scenarios: {  in the first, the external field is} wholly inside the compacton {  (enabling, in this case, to consider the external field as a magnetic impurity),} and the second regards the compacton entirely {within the external field.} Despite both situations, the Bogomol'nyi bound, self-dual equations, and the BPS potential remain the same in both scenarios. {  For an external field} that satisfies the constraints given in Eq. (\ref{Imp0}) or Eq. (\ref{Imp2}), the requirement of the compacton possessing finite total energy leads to imposing that the energy density is null at the compacton border. This condition changes the boundary conditions satisfied by the magnetic field and the potential, as verified when comparing Eqs. (\ref{bcBQU}) and (\ref{bcBUext}).

%%%%%%%%%%%%%%%%%%%%%%%%%%%%%%%%%%%%%%%%%%%%%%%%%%%%%%%%%%%%%%%%%%%%%

To deepen the study, we have selected {two different functions to define the external magnetic field.} The first has a finite size (step-type function), and the second extends along the radial axis (Gaussian function). The analysis of the step-type function (Sec. \ref{stepI}) reveals that the amplitude $c$ has a lower limit for the internal scenario and an upper limit for the extended scenario. Furthermore, the gauge field profile leaves its monotonic behavior, which now can present a maximum (in the internal case) and a minimum (in the extended case), implying a magnetic field flipping (top panels in Fig. \ref{FigB}), a significant feature indicating the magnetic flux local inversion phenomenon. In addition, a detailed analysis of the magnetic flux dependence on the electromagnetic coupling constant has shown that $g$ has restricted values depending on whether the external function describes {the internal scenario or the extended scenario} (see Fig. \ref{FvsG} and Table \ref{T1}).

{In contrast to the finite-sized {  external field}, the Gaussian function (Sec. \ref{stepII}) generates only compactons described by the extended scenario. In this case, there are no} limitations for the values of the amplitude $c$ nor the values of the electromagnetic coupling constant. On the other hand, some features become {similar to those observed in the finite-sized function,} for example, the compacton's radius changes its size with {the external field amplitude, playing an analogous role} to the electromagnetic coupling constant. Besides, the total magnetic flux remains nonquantized, as in the absence of {the external magnetic field.} Still, adjusting the parameters to attain a magnetic flux quantized in units of $2\pi$ becomes possible.  {These findings are of utmost importance, as they provide crucial insights about the behavior of the BPS compact skyrmions depending on the spatial extent of the applied external magnetic field.}
 
We currently investigate BPS structures in generalized or effective versions of the gauged restricted baby Skyrme model, including the emergence of non-compact skyrmions (those whose tails approach the vacuum value following a Gaussian or power-law decay). The exploring encloses nonstandard kinetic terms for both the gauge and Skyrme fields. Furthermore, by incorporating the Chern-Simons term, we are conducting studies to examine the electric and magnetic properties of the gauged baby Skyrme model in the presence of an external magnetic field. We will report the results concerning these studies in future contributions.

%%%%%%%%%%%%%%%%%%%%%%%%%%%%%%%%%%%%%%%%%%%%%%%%
%%%%%%%%%%%%%%%%%%%%%%%%%%%%%%%%%%%%%%%%%%%%%%%%

\begin{acknowledgments}
This study was financed in part by the Coordenaç\~ao de Aperfeiçoamento de
Pessoal de Nível Superior - Brasil (CAPES) - Finance Code 001. We thank also
the Conselho Nacional de Desenvolvimento Cient{\'\i}fico e Tecnol\'ogico
(CNPq), and the Fundaç\~ao de Amparo \`a Pesquisa e ao Desenvolvimento Cient{%
\'\i}fico e Tecnol\'ogico do Maranh\~ao (FAPEMA) (Brazilian Government
agencies). In particular, N. H. G. G. thanks the full support from CAPES; R. C. acknowledges the support from the grants CNPq/312155/2023-9, FAPEMA/UNIVERSAL-00812/19, and FAPEMA/APP-12299/22;  A. C. S. thank the grants CAPES/88882.315461/2019-01 and CNPq/150402/2023-6.
\end{acknowledgments}

%%%%%%%%%%%%%%%%%%%%%%%%%%%%%%%%%%%%%%%%%%%%%%%%
%%%%%%%%%%%%%%%%%%%%%%%%%%%%%%%%%%%%%%%%%%%%%%%%

\end{document}